\documentclass[a4paper,11pt]{article}
\pdfoutput=1 

\usepackage[T1]{fontenc} 

\usepackage{amsmath}
\usepackage{bm} 
\usepackage{amssymb}
\usepackage{graphicx}
\usepackage{color}
\usepackage[usenames,dvipsnames,svgnames,table]{xcolor}
\usepackage{soul}
\usepackage{ulem}
\usepackage{physics}

\usepackage{caption}
\usepackage{subcaption}

\usepackage{booktabs}
\usepackage{multirow}
\usepackage{makecell}

\usepackage{chngcntr}
\counterwithout{equation}{section} 

\usepackage[section]{placeins} 

\usepackage{jcappub} 
                     
\title{Dark matter-baryon scattering effects on temperature perturbations and implications for cosmic dawn}

\author[a,b]{Kathleen Short,}
\author[c]{Jos\' e Luis Bernal,}
\author[d]{Kimberly K.~Boddy,}
\author[e]{Vera Gluscevic}
\author[f]{and Licia Verde}

\affiliation[a]{ICC, University of Barcelona, IEEC-UB, Mart\' i i Franqu\` es, 1, E08028
Barcelona, Spain}
\affiliation[b]{Dept. de F\' isica Qu\` antica i Astrof\' isica, Universitat de Barcelona, Mart\' i i Franqu\` es 1, E08028 Barcelona,
Spain}
\affiliation[c]{William H. Miller III Department of Physics and Astronomy, Johns Hopkins University,  3400 North Charles Street, Baltimore, MD 21218, United States}
\affiliation[d]{Department of Physics, The University of Texas at Austin, Austin, TX, 78712, USA}
\affiliation[e]{Department of Physics and Astronomy, University of Southern California, Los Angeles, CA, 90089, USA}
\affiliation[f]{ICREA, Pg. Llu\' is Companys 23, 08010 Barcelona, Spain}

\emailAdd{katie.short@icc.ub.edu}
\emailAdd{jbernal2@jhu.edu}
\emailAdd{kboddy@physics.utexas.edu}
\emailAdd{vera.gluscevic@usc.edu}
\emailAdd{liciaverde@icc.ub.edu}

\abstract{The nature of dark matter remains unknown, but upcoming measurements probing the high-redshift Universe may provide invaluable insight. In the presence of dark matter-baryon scattering, the suppression in the matter power spectrum and the colder mean gas temperature are expected to modify the evolution of cosmic dawn and reionization. However, the contributions from such interactions to the baryon and dark matter temperature perturbations have been neglected thus far. In this work, we derive these contributions, evolve the cosmological perturbations until the end of the dark ages and show that they may have a significant impact in the beginning of cosmic dawn. In particular, we find that the amplitude of the temperature power spectrum at large scales can change by up to 1--2 orders of magnitude and that the matter power spectrum is further suppressed with respect to $\Lambda$CDM by $5$--$10\%$ at $k\sim 200\, {\rm Mpc^{-1}}$ compared to the computation ignoring these contributions for scattering cross sections at current CMB limits. As a case example, we also compute the HI power spectrum from the dark ages, finding significant differences due to the changes in the temperature and ionization fraction power spectra. We argue that these new contributions must be included in studies of this dark matter model relying on cosmic dawn and reionization observables.}

\begin{document}
\maketitle
\flushbottom

\section{Introduction}\label{sec:intro}
Cosmological observables provide very tight constraints on the abundance of matter in the Universe and in particular on the abundance of the dark matter (DM)~\cite{Planck:2018vyg,eBOSS:2020yzd,DES:2021wwk,Heymans:2020gsg,Brout:2022vxf}. 
Understanding the microscopic nature of DM by exploring phenomenology beyond the known gravitational interactions is one of the primary goals in cosmology and astroparticle physics. Regarding particle DM, different strategies include direct detection~\cite{Billard:2021uyg} and collider~\cite{Boveia:2018yeb} experiments, but cosmological and astrophysical searches often provide sensitivities at different masses ranges~\cite{Boddy:2022knd}. However, despite decades of searches yielding progressively stronger bounds on the nature of the DM, the solution remains elusive. 

Many well-motivated particle candidates for DM, including weakly-interacting massive particle (WIMP) models~\cite{Bertone_2005,Roszkowski:2017nbc}, predict elastic scattering between the DM and baryons which has cosmological consequences. While direct detection experiments are highly sensitive to DM-nucleon interactions for DM masses larger than a few GeV, cosmological probes cover complementary regions of parameter space and are particularly sensitive to scattering of sub-GeV DM with baryons~\cite{Chen:2002yh, Sigurdson:2004zp, Dvorkin:2013cea, Gluscevic:2017ywp, Boddy:2018kfv,Boddy:2018wzy, Xu:2018efh, Slatyer:2018aqg, Li:2018zdm, Nadler:2019zrb, DES:2020fxi, Nguyen:2021cnb,Buen-Abad:2021mvc, Maamari:2020aqz, Bechtol:2019acd, Ooba:2019erm, Ali-Haimoud:2015pwa, Ali-Haimoud:2021lka,Rogers:2021byl}. In a cosmological context, the interaction between DM and baryons leads to transfer of heat and momentum between the two fluids which can affect a range of physical scales and leave imprints on cosmological observables throughout cosmic history. A key signature of the interactions is a suppression of power in cosmological perturbations on small physical scales.

In cosmology, DM-baryon scattering is often characterised by a power law dependence of the momentum-transfer cross section on the relative particle velocity, $\sigma(v)=\sigma_0 v^n$, with a power law index $n$ and an unknown coefficient $\sigma_0$ that quantifies the strength of the interaction at hand. If the dependence of the scattering cross section on the relative velocity is steep enough ($n<-2$), its effect is larger later on in cosmic history, thus impacting late-time observables more than the cosmic microwave background (CMB) anisotropy. Moreover, the impact of the interactions on the baryon temperature and matter clustering at small scales is expected to affect the formation of the first stars. 

Upcoming cosmological probes, in particular line-intensity mapping, will provide new avenues to investigate the nature of DM. While post-reionization measurements are promising in this regard (see e.g.,~\cite{Bauer:2020zsj,Bernal:2020lkd}), the HI intensity mapping signal from cosmic dawn is expected to be especially sensitive to the DM microphysics. In general, any model with clustering suppression at small scales delays the collapse of the first halos and therefore the formation of the first stars, affecting the time evolution of cosmic dawn and reionization~\cite{Das:2017nub,Escudero:2018thh,Safarzadeh:2018hhg,Lopez-Honorez:2018ipk,Munoz:2019hjh,Munoz:2020mue,Jones:2021mrs, Giri:2022nxq, Sarkar:2022dvl}. Furthermore, since the HI signal is highly sensitive to the gas temperature through the HI spin temperature, any modification to the gas temperature (either due to energy injection or heat transfer) is expected to greatly modify the HI signal~\cite{Evoli:2014pva,Fialkov:2018xre,Lidz:2018fqo,Munoz:2018pzp,Munoz:2018jwq,Kovetz:2018zan,Mena:2019nhm, Short:2019twc, Bernal:2017nec}. This has motivated studies of the signatures of DM-baryon scattering in the HI intensity mapping from the dark ages and cosmic dawn~\cite{Tashiro:2014tsa, Munoz:2015bca,Kovetz:2018zan,Slatyer:2018aqg, Munoz:2018pzp, Munoz:2018jwq, Barkana:2018ab, Barkana:2018cct, Berlin:2018sjs}, especially in attempts to explain the anomalous EDGES measurement~\cite{Bowman:2018yin}. 
Experiments such as EDGES~\cite{Bowman:2018yin}, SARAS~\cite{SARAS}, and LEDA~\cite{LEDA}, which measure the sky-averaged brightness temperature of the HI line at cosmic dawn, and HERA~\cite{DeBoer:2016tnn} and SKA~\cite{SKA:2018ckk}, which will provide measurements of its power spectrum, are thus expected to return invaluable insight on the nature of the dark matter. 

However, the direct contribution from DM-baryon scattering to the gas temperature perturbations and the actual dark matter temperature perturbations have been thus far ignored. Temperature perturbations are usually negligible for CMB observations~\cite{Lewis:2007zh} but are critical for signals like the HI intensity mapping, and they may affect the clustering of the first stars. Furthermore, temperature perturbations appear in the sound speed terms of the DM and baryon fluid evolution equations, modifying the clustering at scales below the corresponding Jeans scales. Therefore, we deem the inclusion of these contributions and variables in the evolution equations critical for a detailed assessment of the effects of DM-baryon scattering at low redshift.

In this work, we compute for the first time the effects of DM-baryon elastic scattering in the baryon temperature and ionization fraction perturbations, as well as considering the DM temperature perturbations. We implement this derivation in a modified version of the linear Boltzmann solver \textsc{class}\footnote{\url{https://github.com/lesgourg/class_public}}~\cite{Blas:2011rf}, developed for DM-baryon scattering cosmologies in ref.~\cite{Boddy:2018wzy}, and self-consistently evolve the perturbation equations up to the beginning of cosmic dawn (which we consider to be at $z\sim 30$).\footnote{The specific start of cosmic dawn depends on the cosmological and astrophysical model. However, this does not affect our main conclusions.} We show numerical results for the $n=-4$ case with cross section values set to the 95\% confidence level (C.L.) upper limits from CMB analyses when all the dark matter interacts with baryons~\cite{Boddy:2018wzy, Xu:2018efh}. As expected, we find significant changes in the gas temperature and ionization fraction perturbations due to the additional terms sourced by DM-baryon scattering. Furthermore, there is an increased suppression of the baryon and total matter power spectra relative to $\Lambda$CDM when compared to the prediction for DM-baryon scattering models neglecting the newly-derived contributions to the temperature perturbations. We argue that these changes are large enough to have a sizeable impact in the beginning of cosmic dawn and must be taken into account when setting the initial conditions for cosmic dawn. Once the first stars form, the heating of the gas will be dominated by these sources and a dedicated study, beyond the scope of this work, is required to follow the evolution of the gas and dark matter temperature perturbations coherently with the effects of the scattering in the matter clustering. As an example of the impact of the new terms in an observable, we compute the HI power spectrum from the dark ages.

This paper is organised as follows. In section~\ref{sec:boltz}, after reviewing the modified Boltzmann equations for DM-baryon scattering cosmologies, we present the main result of this work: we derive the evolution of the temperature and ionization fraction perturbations sourced by the DM-baryon interactions.
In section~\ref{sec:num_results} we present our numerical results, quantifying the effects on the DM and baryon temperature and ionization fraction perturbations and the consequent effect on the baryon and total matter power spectra. Finally, in section~\ref{sec:4} we discuss the phenomenological impact of the new temperature perturbation contributions on cosmological observables and as a case example, we compute the effects on the HI line-intensity mapping signal during the dark ages. We discuss and conclude in section~\ref{sec:conc}. Hereinafter we focus on the power spectrum at $z=30$, but relevant results at $z=50$ can be found in appendix~\ref{sec:z50}, while the effects of varying the scattering cross section and the fraction of interacting DM are discussed in appendix~\ref{sec:vary_f_and_sigma}.

\section{Evolution of perturbations with DM-baryon scattering}\label{sec:boltz} 
In this section, we derive the contributions to the perturbed DM and baryon temperatures and ionization fraction arising from the DM-baryon collision term, and incorporate these contributions into the modified Boltzmann equations for DM-baryon interacting cosmologies.
We consider DM-baryon scattering of the form $\sigma(v) = \sigma_0 v^{n}$, where $v$ is the relative velocity between scattering particles. 

Perturbations in the ionization fraction during recombination and the dark ages affect the evolution of the baryon temperature and density perturbations, which has a direct impact on the HI signal and structure formation.
Here we follow the nomenclature introduced by ref.~\cite{Lewis:2007zh} and refer to this effect as ``perturbed recombination''.

\subsection{Modified Boltzmann equations}\label{sec:MBe}
We review the modified Boltzmann equations accounting for scattering between the DM and baryons, following the treatment presented in ref.~\cite{Boddy:2018wzy}. Note that here we work in the Newtonian gauge, but the equations in the synchronous gauge are identical up to terms involving the metric perturbations. The evolution equations for the DM and baryon density perturbations, $\delta_\chi$ and $\delta_b$, and velocity divergences, $\theta_\chi$ and $\theta_b$, respectively, are given by
\begin{align}
\begin{split}\label{eq:fluid}
\dot{\delta_\chi} ={}& -\theta_\chi + 3\dot{\phi} \\
\dot{\delta_b} ={}& -\theta_b + 3\dot{\phi} \\
\dot{\theta_\chi} ={}& -\frac{\dot{a}}{a}\theta_\chi + c^2_{\chi} k^2 \left(\delta_\chi + \delta_{T_\chi}\right)+ k^2\psi + \widetilde{R}_{\chi}\left(\theta_{b} - \theta_\chi \right) \\
\dot{\theta_b} ={}& -\frac{\dot{a}}{a}\theta_b + c^2_b k^2 \left(\delta_b + \delta_{T_b}\right) + k^2\psi + R_{\gamma}\left(\theta_{\gamma} - \theta_b \right)
 + \frac{\rho_\chi}{\rho_b}\widetilde{R}_{\chi}\left(\theta_{\chi} - \theta_b \right),
\end{split}
\end{align}
where $k$ is the wave number of a given Fourier mode; $a$ is the scale factor; $\phi$ and $\psi$ are the Newtonian scalar metric potentials; $c_{\chi}$ and $c_b$ are the DM and baryon sound speeds, respectively; $\rho_\chi$ and $\rho_b$ are their energy densities; $R_\gamma$ is the momentum-transfer rate for baryon-photon coupling due to Compton scattering; and $\widetilde{R}_{\chi}$ is the (modified) momentum-transfer rate arising from the DM-baryon interactions. The overdot denotes a derivative with respect to conformal time. We have added the temperature perturbations $\delta_{T_\chi}$ and $\delta_{T_b}$ ($\delta_{T_{\chi,b}}\equiv \delta T_{\chi,b}/\bar{T}_{\chi,b}$, where $\bar{T}_{\chi,b}$ denotes the mean temperatures) of the DM and baryon gas to the sound speed terms in the momentum equations, as is done for the baryons in refs.~\cite{Naoz:2005pd,Tseliakhovich:2010yw}. Previously the DM temperature perturbations $\delta_{T_\chi}$ have been neglected entirely; in this work, as will become clear in the following, we include it for the first time. As we will see below, accounting for the DM temperature perturbations will not only change the gas temperature power spectrum but also the growth of matter overdensities at small scales, due to this additional contribution to the DM sound speed.

At early times ($z>10^4$), the DM-baryon relative bulk velocity is small compared to the relative thermal velocity dispersion, and the momentum-transfer coefficient $R_\chi$ is approximately independent of the velocity divergences such that
\begin{equation}\label{eq:Rx_orig}
R_{\chi} = a \rho_b  \frac{Y_H \sigma_o}{m_\chi + m_b} \mathcal{N}_n \bar{v}_{\rm{th}}^{(n+1)},
\end{equation}
where $\mathcal{N}_n\equiv 2^{(5+n)/2} \Gamma(3+ n/2)/(3 \sqrt{2})$, $\bar{v}_{\rm{th}}^{2} = T_{\chi}/m_{\chi} + T_b/m_b$ is the relative thermal velocity dispersion squared, $T_\chi \,(m_\chi)$ and $T_b \,(m_b)$ are the DM and baryon temperatures (masses), respectively,  $Y_H$ is the mass fraction of hydrogen, and we neglect helium scattering. This approximation breaks down at lower redshifts for $n\leq -2$ as the relative bulk velocity becomes non-negligible, resulting in non-linear Boltzmann equations and mode-coupling. In order to restore linearity to the evolution equations, we follow the prescription of ref.~\cite{Boddy:2018wzy} in which a modified momentum-transfer rate is introduced to include the effects of mode mixing.  
Defining
\begin{align}
\begin{split}\label{eq:Vrms_Vflow}
V^2_{\rm{flow}} \equiv{}& \int_{0}^{k} \, \frac{dk'}{k'} \Delta^2_{\zeta} \left [ \frac{\theta_b(k', z) - \theta_{\chi}(k', z)}{k'} \right ]^2 \\[5pt]
V^2_{\rm{RMS}} \equiv{}& \int_{k}^{\infty} \, \frac{dk'}{k'} \Delta^2_{\zeta} \left [ \frac{\theta_b(k', z) - \theta_{\chi}(k', z)}{k'} \right ]^2,
\end{split}
\end{align}
where $\Delta^2_{\zeta}$ is the primordial curvature perturbation variance per $\ln k$, the modified momentum-transfer rate is given by
\begin{equation}
\widetilde{R}_{\chi} = R_\chi \left [ 1 + \frac{V^2_{\rm{RMS}}(k, z)/3}{\bar{v}_{\rm{th}}^{2}} \right]^{(n+1)/2} \times {}_1F_1\left(-\frac{n+1}{2}, \frac{5}{2}, -\frac{V^2_{\rm{flow}}(k,z)}{2 [\bar{v}^2_{\rm{th}} + V^2_{\rm{RMS}}(k, z)/3 ]} \right)\,,
\end{equation}
where ${}_1F_1$ is the confluent hypergeometric function of the first kind. 
In the limit where $ \langle V^2_{\chi b} \rangle \equiv V^2_{\rm{flow}} + V^2_{\rm{RMS}} \ll \bar{v}^2_{\rm{th}}$,\footnote{$\langle V^2_{\chi b} \rangle$ is the full variance of the relative bulk velocity, integrated over all $k$.} we recover $\widetilde{R}_{\chi} = R_\chi$, as desired. 

Before recombination, supersonic relative velocities between the DM and baryons are generated, leading to a root-mean-square velocity of $\sim 30$ km s$^{-1}$ at the time of recombination. This supersonic relative velocity permits the baryons to freely stream out of the DM potential wells, suppressing density perturbations at scales below the characteristic advection scale over a Hubble time at the time of decoupling ($k\sim 40 \, \rm Mpc^{-1}$)~\cite{Tseliakhovich:2010yw,Tseliakhovich:2010bj,Dalal:2010yt,Fialkov:2011iw}. The relative velocity is coherent over scales smaller than $k\sim 0.3 \, \rm Mpc^{-1}$, which allows the use of moving-background perturbation theory to account for a formally non-linear term, by perturbing around a local background value of the relative bulk velocity within coherent patches. 
Following ref.~\cite{Tseliakhovich:2010bj}, the evolution of small-scale perturbations after recombination in the presence of a local relative bulk velocity $\boldsymbol{v}_{\chi b}^{\rm bg}$ can be written to first order as\footnote{Note that, contrary to ref.~\cite{Tseliakhovich:2010bj}, we use derivatives with respect to conformal time and use the standard definition of $\theta$ as the velocity divergence with respect to comoving space.}
\begin{align}
\begin{split}\label{eq:mbpt_eq}
\dot{\delta}_\chi ={}& i(\vb*{v}^{\,\rm{bg}}_{\chi b} \cdot \vb*{k})\, \delta_\chi - \theta_\chi\, \\
\dot{\theta}_\chi ={}& i(\vb*{v}^{\,\rm{bg}}_{\chi b} \cdot \vb*{k})\theta_\chi - \frac{\dot{a}}{a}\theta_\chi + k^2\phi + c^2_\chi k^2 (\delta_\chi+\delta_{T_{\chi}}) + \widetilde{R}_\chi(\theta_b - \theta_\chi)\, \\
\dot{\delta}_b ={}&  - \theta_b\, \\
\dot{\theta}_b ={}&  - \frac{\dot{a}}{a}\theta_b+ k^2\phi + c^2_b k^2 (\delta_b + \delta_{T_{b}})+ \frac{\rho_\chi}{\rho_b} \widetilde{R}_\chi(\theta_\chi - \theta_b)\, \\
\textrm{with} \quad  \frac{k^2}{a^2}\phi ={}& -\frac{3}{2}\frac{H^{2}_0}{a^3} (\Omega^0_b \delta_b + \Omega^0_\chi \delta_\chi)\,.
\end{split}
\end{align}
In this regime, we are considering scales much smaller than the horizon and times after recombination, and so we neglect the relativistic contributions to the evolution equations and the baryon-photon coupling term. We assume that $\vb*{v}^{\,\rm{bg}}_{\chi b}$ follows an isotropic Gaussian distribution with variance given by $\langle V^2_{\chi b} \rangle=V^2_{\rm{flow}}+V^2_{\rm{RMS}}$ and place ourselves in the local baryon rest frame.

The small-scale patches with coherent relative velocities are correlated as a relic of the correlation of the relative velocities after recombination. Therefore, the small-scale fluctuations modulated by the relative velocities also affect the large-scale fluctuations through uncorrelated, additive quadratic terms~\cite{Tseliakhovich:2010yw,Dalal:2010yt}. The quadratic terms at small scales are averaged over regions of the size of the coherence scale of the relative velocity and contribute to large-scale perturbations, following the correlation function of the local relative velocities. 

\subsection{Temperature and ionization fraction perturbations}\label{sec:temp_perts}
To derive the evolution equations for the DM and baryon temperature perturbations, we begin from the first law of thermodynamics
\begin{equation}\label{eq:1stlaw}
\frac{3}{2} dT_{\chi,b} - T_{\chi,b} d \log \rho_{\chi,b} = dQ_{\chi,b},
\end{equation}
where $dQ_{\chi,b}$ is the heating rate per particle. In the absence of DM-baryon interactions, the only relevant source of heating is the standard Compton collision heating rate for baryons scattering with CMB photons, given by 
\begin{equation}\label{eq:Q_C}
\dot{Q}_C = 3 \frac{\mu_b}{m_e} R_\gamma (T_\gamma - T_b) = \frac{4a\sigma_T \rho_\gamma x_e}{m_e (1 + x_{\rm He} + x_e)} (T_\gamma - T_b)
\end{equation}
where $\mu_b = m_H(1 + 4x_{\rm He})/(1 + x_{\rm He} + x_e)$ is the mean molecular weight of the baryons, $m_e$ is the electron mass, $T_\gamma$ is the photon temperature, $x_e=n_e/n_{\rm H}$ is the free electron fraction, $x_{\rm He}=n_{\rm He}/n_{\rm H}$ is the constant helium to hydrogen number density ratio, $\rho_\gamma=a_rT_{\gamma}^4$ is the photon energy density with radiation constant $a_r$, and $\sigma_T$ is the Thompson cross section. 

\begin{figure}[tbp]
\centering
\includegraphics{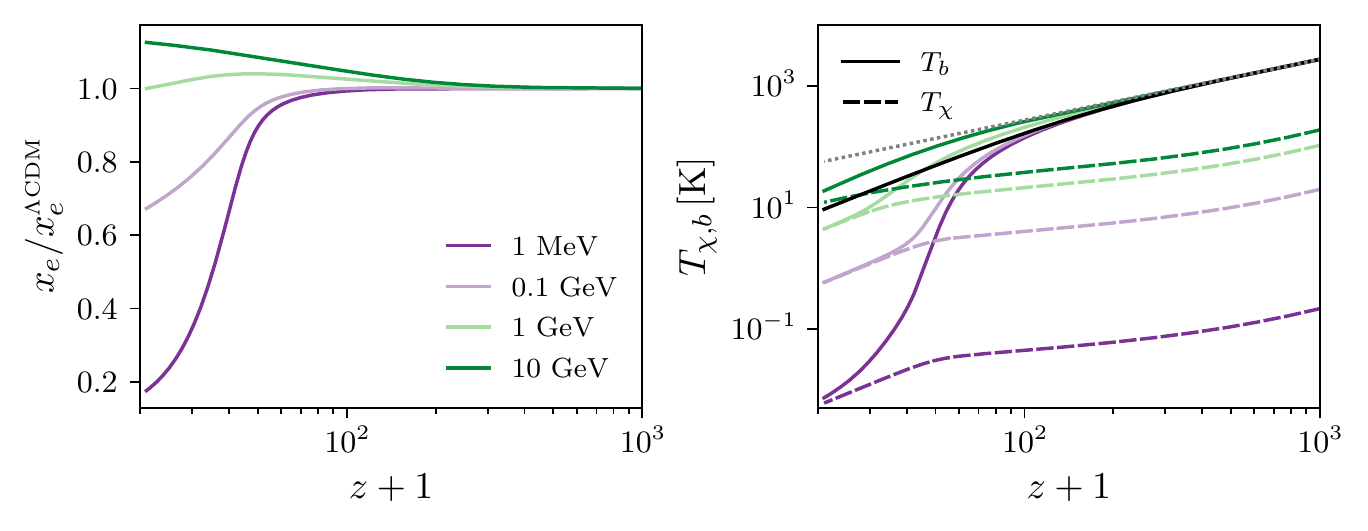}
\centering{\caption{\label{fig:Vxb_Temp} Evolution of the mean free-electron fraction over the $\Lambda$CDM prediction (left) and evolution of the mean baryon and DM temperatures (right) as a function of redshift for several DM masses for $n=-4$. The $\Lambda$CDM mean baryon temperature is shown in black, and the CMB temperature is shown in dotted grey. For each DM mass, the coefficient of the momentum-transfer cross section $\sigma_0$ can be found in table~\ref{tab:sigma0}, consistent with $95\%$ C.L. upper limits derived in refs.~\cite{Boddy:2018wzy, Xu:2018efh}.
}}
\end{figure}

In our case, there is an additional contribution sourced by DM-baryon interactions $\dot{Q}^{\rm iDM}$, for both the baryons and DM. Scattering of the baryons with the colder DM leads to heat exchange between the two species, cooling down the baryons and heating the DM. Moreover, the non-negligible relative bulk velocity induces a drag force between the two fluids which generates an additional temperature-independent heating term that dominates over the baryon cooling at high redshift for masses $m_\chi \gtrsim 1$ GeV~\cite{Munoz:2015bca}.
The heat exchange rates due to DM-baryon interactions, accounting for a non-negligible relative bulk velocity, were generalised for all $n>-5$ in ref.~\cite{Boddy:2018wzy}; in what follows, we substitute $V^2_{\chi b}$ with its averaged value $\langle V^2_{\chi b} \rangle$ to maintain linearity, according to the prescription set out in that work. Let us define
\begin{align}
\begin{split}\label{eq:coeffs_full}
\Gamma_\chi \equiv{}&  2 \frac{\bar{\rho}_b }{\rho_b} R'_{\chi} \left[  {}_1F_1 \left(-\frac{n+3}{2}, \frac{3}{2}, -\frac{\langle V^2_{\chi b} \rangle}{2\bar{v}^2_{\rm{th}}} \right) - \frac{\langle V^2_{\chi b} \rangle}{3\bar{v}^2_{\rm{th}}} {}_1F_1\left(-\frac{n+1}{2}, \frac{5}{2}, -\frac{\langle V^2_{\chi b} \rangle}{2\bar{v}^2_{\rm{th}}}\right) \right]\,, \\[10pt]
\Gamma_b\equiv{}& \frac{2\mu_b}{m_\chi} \frac{\bar{\rho}_\chi }{\rho_b} R'_\chi \left[  {}_1F_1 \left(-\frac{n+3}{2}, \frac{3}{2}, -\frac{\langle V^2_{\chi b} \rangle}{2\bar{v}^2_{\rm{th}}} \right) - \frac{\langle V^2_{\chi b} \rangle}{3\bar{v}^2_{\rm{th}}} {}_1F_1\left(-\frac{n+1}{2}, \frac{5}{2}, -\frac{\langle V^2_{\chi b} \rangle}{2\bar{v}^2_{\rm{th}}}\right) \right]\,, \\[10pt]
\mathcal{H}_\chi\equiv{}& 2 \frac{\bar{\rho}_b }{\rho_b} R'_{\chi}\frac{m_b}{3} \langle V^2_{\chi b} \rangle  {}_1F_1\left(-\frac{n+1}{2}, \frac{5}{2}, -\frac{\langle V^2_{\chi b} \rangle}{2\bar{v}^2_{th}}\right)\,, \\[10pt]
\mathcal{H}_b\equiv{}&  \frac{2\mu_b}{m_\chi} \frac{\bar{\rho}_\chi }{\rho_b} R'_\chi  \frac{m_\chi}{3} \langle V^2_{\chi b} \rangle  {}_1F_1\left(-\frac{n+1}{2}, \frac{5}{2}, -\frac{\langle V^2_{\chi b} \rangle}{2\bar{v}^2_{th}}\right),
\end{split} 
\end{align}
where the heat-transfer rate coefficient is given by $R'_\chi = R_\chi m_\chi/(m_\chi + m_b)$.
The heating rates due to DM-baryon interactions may then be written as
\begin{align}
\begin{split}\label{eq:Q_rates}
\dot{Q}^{\rm iDM}_\chi ={} \frac{3}{2} \Gamma_\chi \frac{\rho_b}{\bar{\rho}_b}(T_b - T_\chi) + \mathcal{H}_\chi \frac{\rho_b}{\bar{\rho}_b}\,,\qquad 
 \dot{Q}^{\rm iDM}_b  ={} \frac{3}{2} \Gamma_b \frac{\rho_\chi}{\bar{\rho}_\chi}(T_\chi - T_b ) + \mathcal{H}_b \frac{\rho_\chi}{\bar{\rho}_\chi}\,.
\end{split} 
\end{align}
Combining eqs.~\eqref{eq:1stlaw}, \eqref{eq:Q_C} and~\eqref{eq:Q_rates}, we obtain the full evolution equations for the DM and baryon temperatures
\begin{align}
\begin{split}\label{eq:dtemp_full}
\dot{T_\chi} - \frac{2}{3}\frac{\dot{\rho}_\chi}{\rho_\chi} T_\chi ={}& \Gamma_\chi \frac{\rho_b}{\bar{\rho}_b}(T_b - T_\chi) + \mathcal{H}_\chi \frac{\rho_b}{\bar{\rho}_b}\,, \\[10pt]
 \dot{T_b}  - \frac{2}{3}\frac{\dot{\rho}_b}{\rho_b} T_b  ={}& \Gamma_b \frac{\rho_\chi}{\bar{\rho}_\chi}(T_\chi - T_b ) + \mathcal{H}_b \frac{\rho_\chi}{\bar{\rho}_\chi} + \frac{2\mu_b}{m_e} R_\gamma \left(T_\gamma-T_b\right). 
\end{split}
\end{align}
Expanding eq.~\eqref{eq:dtemp_full} and linearising using $\dot{\rho}_{i}/\rho_{i} = -3\dot{a}/a + \dot{\delta}_{i}$ for species $i$, we find the homogeneous solution gives the evolution of the background temperatures 
\begin{align}
\begin{split}\label{eq:dtemp_bkg}
\dot{\bar{T}}_\chi + 2\frac{\dot{a}}{a}\bar{T}_\chi ={}& \Gamma_\chi (\bar{T}_b - \bar{T}_\chi) + \mathcal{H}_\chi\,, \\[10pt]
\dot{\bar{T}}_b + 2\frac{\dot{a}}{a}\bar{T}_b ={}& \Gamma_b (\bar{T}_\chi - \bar{T}_b) + \mathcal{H}_b + \frac{2\mu_b}{m_e} R_\gamma \left(\bar{T}_\gamma-\bar{T}_b\right)\,,
\end{split}
\end{align}
which matches the results found in previous work~\cite{Munoz:2015bca, Boddy:2018wzy}. In figure~\ref{fig:Vxb_Temp} we show the evolution of the background temperatures as function of redshift for different DM masses and $n=-4$ when setting $\sigma_0$ to current 95\% C.L. upper limits from CMB power spectrum analyses as reported in table~\ref{tab:sigma0}. 

At first order, the perturbations $\delta_{T_i}=(T_i-\bar{T}_i)/\bar{T}_i$ to the DM and gas temperatures then evolve according to
\begin{align}
\dot{\delta}_{T_\chi} - \frac{2}{3} \dot{\delta}_\chi ={}& \Gamma_\chi  \left[ \left( \frac{\bar{T}_b - \bar{T}_\chi}{\bar{T}_\chi}\right)\delta_b + \frac{\bar{T}_b}{\bar{T}_\chi}(\delta_{T_b} - \delta_{T_\chi}) \right] +  \mathcal{H}_\chi \left(\frac{\delta_b - \delta_{T_\chi}}{\bar{T}_\chi}\right) \label{eq:Tdm_1stO} \\[10pt] 
\begin{split}\label{eq:Tgas_1stO}
\dot{\delta}_{T_b} - \frac{2}{3} \dot{\delta}_b ={}&  \Gamma_b \left[ \left( \frac{\bar{T}_\chi - \bar{T}_b}{\bar{T}_b}\right)\delta_\chi + \frac{\bar{T}_\chi}{\bar{T}_b}(\delta_{T_\chi} - \delta_{T_b}) \right] + \mathcal{H}_b \left (\frac{\delta_\chi - \delta_{T_b}}{\bar{T}_b}\right )\, \\[10pt]
&+ \frac{2\mu_b}{m_e} R_\gamma \left[\left(\frac{\bar{T}_\gamma}{\bar{T}_b} -1  \right) \left \{\delta_{x_e} - \frac{\bar{x}_e \delta_{x_e}}{1 + x_{\rm{He}}+\bar{x}_e} + \delta_\gamma \right \} - \frac{\bar{T}_\gamma}{\bar{T}_b} ( \delta_{T_b} - \delta_{T_\gamma})\right]\,,
\end{split}
\end{align}
where $\delta_{T_\gamma}$ is the photon temperature fluctuation, $\delta_\gamma = 4\delta_{T_\gamma}$ is the photon density fluctuation, $\delta_{x_e} \equiv \delta x_e / \bar{x}_e $ is free electron fraction fluctuation, and helium fraction perturbations are neglected. In both equations the first two terms correspond to the contributions from the DM-baryon scattering,
while the contributions from Compton heating are encoded in the second line of the baryon temperature perturbation equation. A detailed derivation of the fluctuations sourced by the standard Compton heating term can be found in e.g., refs.~\cite{Lewis:2007zh,Naoz:2005pd,Senatore:2008vi}. Eqs.~\eqref{eq:Tdm_1stO} and~\eqref{eq:Tgas_1stO} are the main theoretical result of this work, and we study the effects of including the temperature perturbations in the following sections.

At small scales and after recombination, where the supersonic relative velocities between DM and baryons are coherent and the growth of perturbations is affected, we solve the system using the moving-background perturbation theory as described in eq.~\eqref{eq:mbpt_eq}. Temperature (and ionization fraction, see below) perturbations are affected by the coherent bulk velocities indirectly through the DM and baryon perturbations, hence we need to take this into account. Furthermore, we neglect the photon temperature perturbations in this regime as they are negligible.

\subsubsection{Ionization fraction fluctuations}
The evolution of the ionization fraction perturbations is modified by DM-baryon interactions indirectly through the dependence on the gas temperature fluctuations. After recombination, the free-electron fraction obeys~\cite{Lewis:2007zh}
\begin{equation}\label{eq:x_e}
\dot{x}_e \approx - a \alpha_{\rm{B}} n_H x^2_e,
\end{equation}
where $\alpha_B$ is the case-B recombination coefficient used in \textsc{recfast}~\cite{Seager:1999bc},
\begin{equation}
    \alpha_B = F \frac{a (T_b/10^4 \, \mathrm{K} )^b}{1 + c(T_b/10^4 \, \mathrm{K})^d} \, \mathrm{m}^3 \mathrm{s}^{-1}\,,
\end{equation}
with $a = 4.309 \times 10^{-19}$, $b = -0.6166$, $c = 0.6703$, $d = 0.5300$ and $F= 1.125$ is a fudge factor to reproduce the result of a multi-level atom calculation~\cite{Rubino-Martin:2009frf,Lee:2020obi}. The linear ionization fraction perturbations then evolve as
\begin{equation}\label{eq:delta_xe}
\dot{\delta}_{x_e} = \frac{\dot{\bar{x}}_e}{\bar{x}_e} (\delta_{x_e} + \delta_b + \delta_{\alpha_{\rm{B}}}),
\end{equation}
where $\delta_{n_H} \approx \delta_b$ up to very small corrections, and 
\begin{equation}\label{eq:delta_alpha}
\delta_{\alpha_{\rm{B}}} = \frac{\partial \ln \mathcal{\alpha}_{\rm{B}}}{\partial \ln T_b}\delta_{T_b}\,.
\end{equation}

\subsubsection{Large-scale enhancement of temperature and ionization perturbations}
At small scales, the relative velocities only affect the temperature and ionization perturbations indirectly through their effect on the matter. Over large scales, the small-scale relative velocities have no dynamical effect on the growth of overdensities, since the non-linear terms are full divergences that integrate to zero. However, the quadratic cooling of temperatures at small scales is non-adiabatic and therefore the temperature and ionization fraction need to be evolved as a coupled system up to second order, which affects the large-scale temperature and ionization perturbations differently~\cite{Ali-Haimoud:2013hpa}.  

Perturbing the temperature evolution equations~\eqref{eq:dtemp_full} to second order -- e.g., $T_{i}=\bar{T}_i(1+\delta_{T_i}^{\rm I}+\delta_{T_i}^{\rm II})$ -- we find
\begin{align}
\dot{\delta}^{\rm{II}}_{T_\chi} ={}& \frac{2}{3} \dot{\delta}_\chi (\delta^{\rm{I}}_{T_\chi} - \delta_\chi) + \Gamma_\chi  \left[ \frac{\bar{T}_b}{\bar{T}_\chi} \left(\delta^{\rm{II}}_{T_b} - \delta^{\rm{II}}_{T_\chi} + \delta^{\rm{I}}_{T_b}\delta_b\right) - \delta^{\rm{I}}_{T_\chi}\delta_b \right] + \mathcal{H}_\chi \frac{ \delta^{\rm{II}}_{T_\chi}}{\bar{T}_\chi} \label{eq:dTx_2} \\[10pt]
\begin{split}
\dot{\delta}^{\rm{II}}_{T_b} ={}& \frac{2}{3} \dot{\delta}_b (\delta^{\rm{I}}_{T_b} - \delta_b) + \Gamma_b  \left[ \frac{\bar{T}_\chi}{\bar{T}_b} \left(\delta^{\rm{II}}_{T_\chi} - \delta^{\rm{II}}_{T_b} + \delta^{\rm{I}}_{T_\chi}\delta_\chi\right) - \delta^{\rm{I}}_{T_b}\delta_\chi \right] + \mathcal{H}_b \frac{ \delta^{\rm{II}}_{T_b}}{\bar{T}_b} \\[10pt]
&\qquad\qquad\quad\ \  + \frac{2\mu_b}{m_e} R_\gamma \left[ \frac{\bar{T}_\gamma - \bar{T}_b}{\bar{T}_b} \delta^{\rm{II}}_{x_e} - \delta^{\rm{I}}_{x_e}\delta^{\rm{I}}_{T_b} - \frac{\bar{T}_\gamma}{\bar{T}_b} \delta^{\rm{II}}_{T_b} \right] \label{eq:dTb_2} ,
\end{split}
\end{align}
where we have neglected photon temperature fluctuations and assumed $1+x_{\rm{He}} + x_e \approx 1+x_{\rm{He}}$ to be locally homogeneous, valid at small (sub-horizon) scales. The second-order perturbations to the ionization fraction $\delta^{\rm{II}}_{x_e}$ evolve according to~\cite{Ali-Haimoud:2013hpa}
\begin{align}\label{eq:dxe_2}
\begin{split}
\dot{\delta}^{\rm{II}}_{x_e} = \frac{\dot{x}_e}{\bar{x}_e} \left(\delta^{\rm{II}}_{x_e} + \frac{d \log \mathcal{A}_B}{d \log T_b} \delta^{\rm{II}}_{T_b} + \delta^{\rm{II}}_{\dot{x}_e}\right),
\end{split}
\end{align}
where $ \delta^{\rm{II}}_{\dot{x}_e}$ represents the part of $\delta \dot{x}_e/\dot{\bar{x}}_e$ quadratic in the perturbations. Each of the source terms in the above equations are previously smoothed over the small volumes with coherent relative velocities.

\section{Numerical results}\label{sec:num_results}
We solve the perturbation equations in different regimes, accounting for all new terms in the temperature perturbations sourced by DM-baryon scattering, and also for the coherent relative velocities at small scales and the consequent quadratic contributions at large scales, as required in each regime. Therefore, we use a different approach in each case. 

We have implemented the new contributions to the perturbed recombination arising from DM-baryon interactions derived in section~\ref{sec:temp_perts} (see eqs.~\eqref{eq:Tdm_1stO}, \eqref{eq:Tgas_1stO} and \eqref{eq:delta_xe}) into a modified version of the linear Boltzmann solver \textsc{class}
that has been developed for DM-baryon scattering models in ref.~\cite{Boddy:2018wzy}. We use this version of \textsc{class} to evolve the linear perturbation equations at all scales before recombination and only large scales ($k\lesssim 10$ Mpc$^{-1}$) after that. We separately solve the evolution equations at small scales post-recombination as function of $k$ and $\boldsymbol{v}_{b\chi}\cdot\boldsymbol{k}$ (the local relative velocity in each patch) and average the results using the probability distribution function of $\boldsymbol{v}_{b\chi}\cdot\boldsymbol{k}$. For this we take the initial conditions from the modified version of \textsc{class} at $z=1010$ and evolve them until $z=20$.\footnote{At $z\lesssim 40\text{--}30$ we expect the first stars to form, triggering cosmic dawn, which will then dominate the temperature perturbations; the effects of DM-baryon scattering on the temperature perturbations during cosmic dawn are beyond the scope of this study and left for future work.} With these results, we follow ref.~\cite{Ali-Haimoud:2013hpa} to compute the quadratic corrections at large scales, using the smoothed perturbations at small scales and the solution of the system in eqs.~\eqref{eq:dTx_2}, \eqref{eq:dTb_2}, and~\eqref{eq:dxe_2}. For $z\lesssim 50$, non-linear effects become important at the several percent level at high-$k$. Due to this, we restrict our $k_{\rm max} = 500\, \rm Mpc^{-1}$ when computing the large-scale modulation of small-scale perturbations at these redshifts, which should result in a conservative estimate.

For all results presented in this work we consider DM-baryon interactions with $n=-4$ velocity dependence and adopt the best-fit \textit{Planck} 2018~\cite{Planck:2018vyg} values for the $\Lambda$CDM cosmological parameters. 
We show results for selected masses in the range 1 MeV--10 GeV, setting $\sigma_0$ for each DM mass to the corresponding 95\% C.L. upper limit found in ref.~\cite{Boddy:2018wzy} where available. The mass-scaling relationship $\sigma_0 \propto (m_\chi + m_H)$ of refs.~\cite{Xu:2018efh, Slatyer:2018aqg} is used to obtain our benchmark 10 GeV limit. For reference these values are reported in table~\ref{tab:sigma0}. As anticipated in the introduction, the perturbed recombination -- even with the new contributions derived in the previous section -- has no significant effect on the CMB power spectra.

\begin{table}[tbp]
\centering
\begin{tabular}{@{}cc@{}}
\toprule
DM mass $m_\chi$ & $\sigma_0 \, [\rm{cm}^2]$ \\ \midrule
1 MeV            & $1.7 \times 10^{-41}$    \\
0.1 GeV          & $1.9\times 10^{-41}$   \\
1 GeV            & $3.5\times 10^{-41}$    \\
10 GeV           & $2.0\times 10^{-40}$    \\ \bottomrule
\end{tabular}
\caption{Benchmark values of the momentum-transfer cross section coefficient $\sigma_0$ for each DM mass used in all of our numerical results. Consistent with the 95\% C.L. upper limits found in refs.~\cite{Boddy:2018wzy,Xu:2018efh, Slatyer:2018aqg}.}
\label{tab:sigma0}
\end{table}

\subsection{Temperature and ionization fraction perturbations}
We first focus on the perturbations of the baryon temperature and ionization fraction. We compare the
results considering DM-baryon scattering with respect to that of $\Lambda$CDM, 
both when including and neglecting the DM-baryon interactions contribution to the perturbed recombination (in the latter case, this amounts to setting $\Gamma_b=\mathcal{H}_b=0$ in eq.~\eqref{eq:Tgas_1stO} and ignoring $\delta_{T_\chi}$). We refer to these two cases as $\delta^{\rm iDM}$ and $\delta^{\rm std}$ and represent them in figures with solid (or dark) and dashed (or light), respectively, unless otherwise stated. 

\begin{figure}[tbp]
\centering
\includegraphics{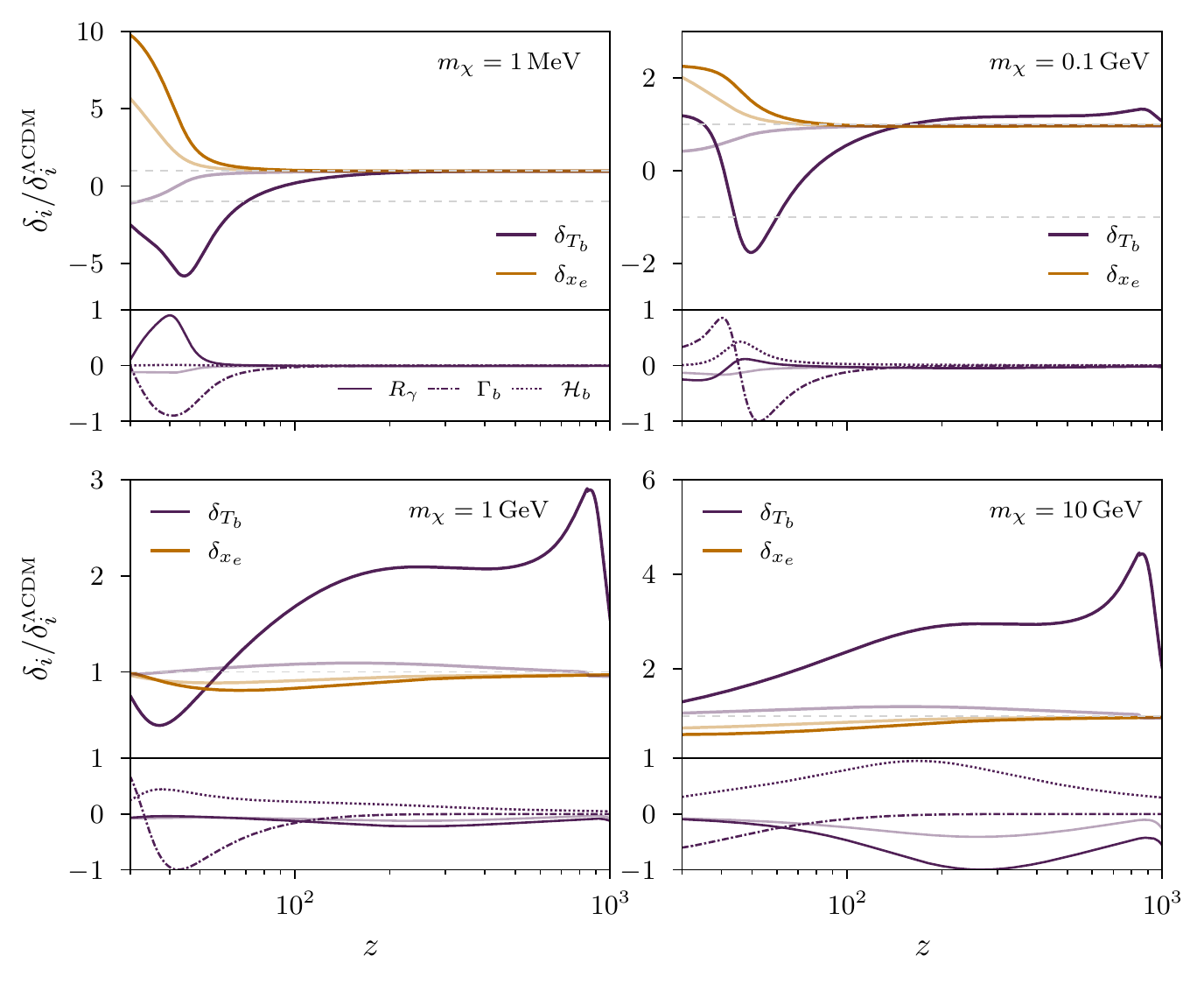}
\centering{\caption{\label{fig:dTgas_xe_k1} The evolution of the gas temperature and ionization fraction fluctuations at $k=1\, \mathrm{Mpc}^{-1}$ as a fraction of the respective fluctuation in $\Lambda$CDM, including (dark) and neglecting (light) the new temperature perturbations sourced by DM-baryon interactions. In the bottom panel of each figure, we show the relative contribution of each of the three terms in eq.~\eqref{eq:Tgas_1stO} (labelled by its coefficient) to the temperature fluctuations, scaled to the range $[-1, 1]$. When neglecting the fluctuations from DM-baryon scattering, there is only a contribution from the term proportional to $R_\gamma$. Note the change of scale in the $y$-axis of each panel. 
}}
\end{figure}

The evolution of the baryon temperature and ionization fraction perturbations as a function of redshift for $k=1$ Mpc$^{-1}$ is shown in figure~\ref{fig:dTgas_xe_k1}. In the bottom panels, we explicitly show the relative contributions of each term in eq.~\eqref{eq:Tgas_1stO} to the evolution of the perturbations, with each term labelled by its rate coefficient respectively: $R_\gamma$,  $\Gamma_b$, or $\mathcal{H}_b$. 
For $m_\chi \gtrsim 1$ GeV, the drag heating term ($\mathcal{H}_b$) dominates at early times, increasing temperature fluctuations and therefore indirectly enhancing the standard Compton heating term through its dependence on $\delta_{T_b}$, analogous to the effect on the mean temperatures. These two terms have opposite impact in the evolution of $\delta_{T_b}$, but their balance results in enhanced temperature perturbations for all redshifts considered. On the other hand, the ionization fraction fluctuations are suppressed relative to $\Lambda$CDM, as recombination is less efficient in the overdensities due to the higher gas temperature, i.e. there are more free electrons. A similar effect -- albeit with smaller amplitude -- is seen for $m_\chi = 1$ GeV, with the exception that the contribution from the cooling term ($\Gamma_b$) becomes important at low-$z$, and $\delta_{T_b}$ is suppressed for $z\lesssim 60$. 
For $m_\chi < 1$ GeV, while the $\mathcal{H}_b$ contribution can enhance temperature perturbations at $z\sim 10^3$, the cooling term dominates at late times; this results first in suppressed temperature fluctuations for $ 70 \lesssim z \lesssim 200$, and later enhanced negative fluctuations $ z \lesssim 70$ compared to $\Lambda$CDM. For $m_\chi = 0.1 $ GeV, the DM and baryon temperatures reach equilibrium at $z\sim 50$, and so the dominant contribution to the cooling term becomes negligible. The remaining terms add up to increase the temperature perturbations, and hence the gas temperature grows again in over-dense regions. For $m_\chi = 1 $ MeV, the DM and baryon temperatures reach equilibrium later, at $z\lesssim 30$. Finally, for these two cases, the lower temperature in overdensities results in an enhancement of the (negative) ionization fraction fluctuations, due to more efficient recombination. In all cases, the interdependence of the three contributions to the temperature perturbations results in a richer redshift evolution of the perturbed temperatures with respect to neglecting the terms sourced by DM-baryon scattering.

\begin{figure}[tbp]
\centering
\includegraphics{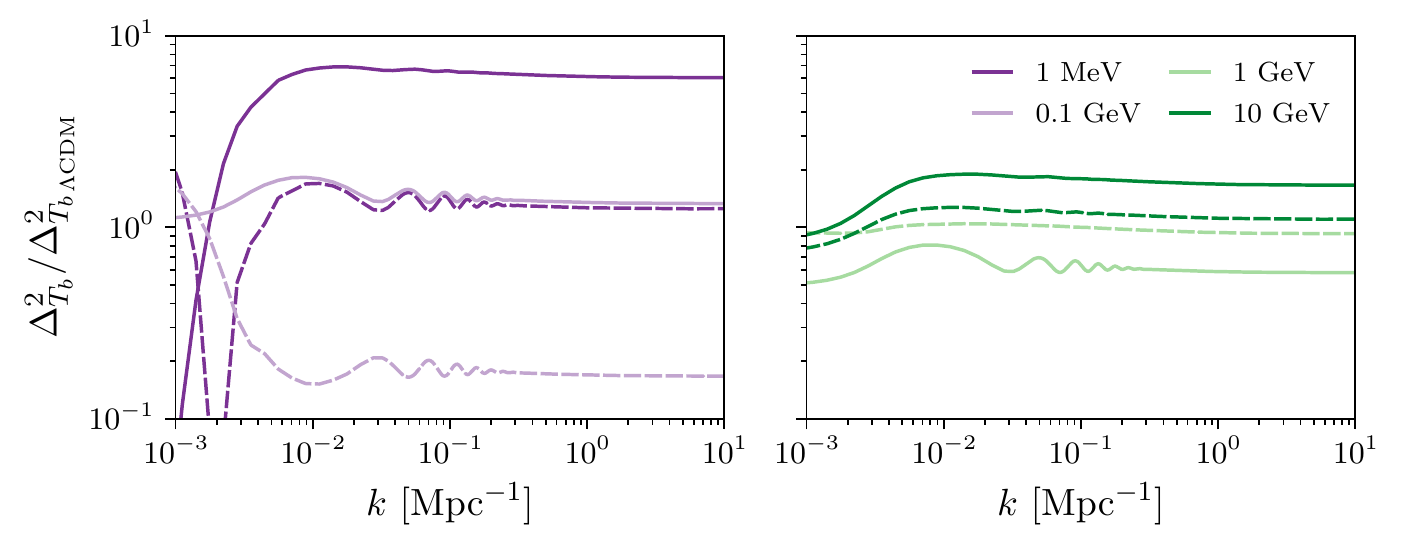}
\centering{\caption{\label{fig:PS_Tgas_z30_z50} Ratio of the linear power spectra (with respect to $\Lambda$CDM) of gas temperature fluctuations at $z=30$ for several DM masses, including (solid) and neglecting (dashed) fluctuations sourced by DM-baryon interactions, shown until $k = 10\, \rm Mpc^{-1}$. At smaller scales, other effects become important, which are considered in section~\ref{sec:small_scales}.}}
\end{figure}

In figure~\ref{fig:PS_Tgas_z30_z50}, we demonstrate how the baryon temperature power spectrum changes due to the DM-baryon interactions and in particular, the importance of the new terms derived in the previous section. We plot the ratio of the baryon temperature power spectrum for different DM masses with respect to the $\Lambda$CDM prediction at $z=30$. The contributions from the DM-baryon scattering to the baryon temperature power spectrum modify the amplitude by a factor of a few up to several tens (even hundreds for $m_\chi = 1$ MeV at $z=50$, see figure~\ref{fig:PS_Tgas_z50}), in a roughly scale-independent manner. Except in the case of 1 GeV, for which temperature fluctuations become suppressed relative to $\Lambda$CDM  for $30 \lesssim z \lesssim 60 $, the power spectrum is always larger than the $\Lambda$CDM prediction at the end of the dark ages ($z \sim 30$--$50$). Moreover, we see that including the DM-baryon scattering contributions to the perturbed recombination can enhance or suppress the baryon acoustic oscillation (BAO) feature in the temperature power spectrum. This is a result of the changes, with respect to considering Compton scattering only, in the terms that dominate the baryon temperature perturbations and their dependence on the density perturbations. 

These results depend on the strength of the interacting cross section, which we explore for a DM mass of 1 GeV in appendix~\ref{sec:vary_f_and_sigma}. Interestingly, the change in amplitude of the gas temperature power spectrum does not necessarily decrease with the scattering cross section; reducing the cross section alters the balance of competing terms on the right-hand side of eq.~\eqref{eq:Tgas_1stO}, which can result in more or less enhancement or suppression of the temperature perturbations depending on the redshift. 

\subsection{Small-scale effects}\label{sec:small_scales}

\begin{figure}[tbp]
\centering
\includegraphics{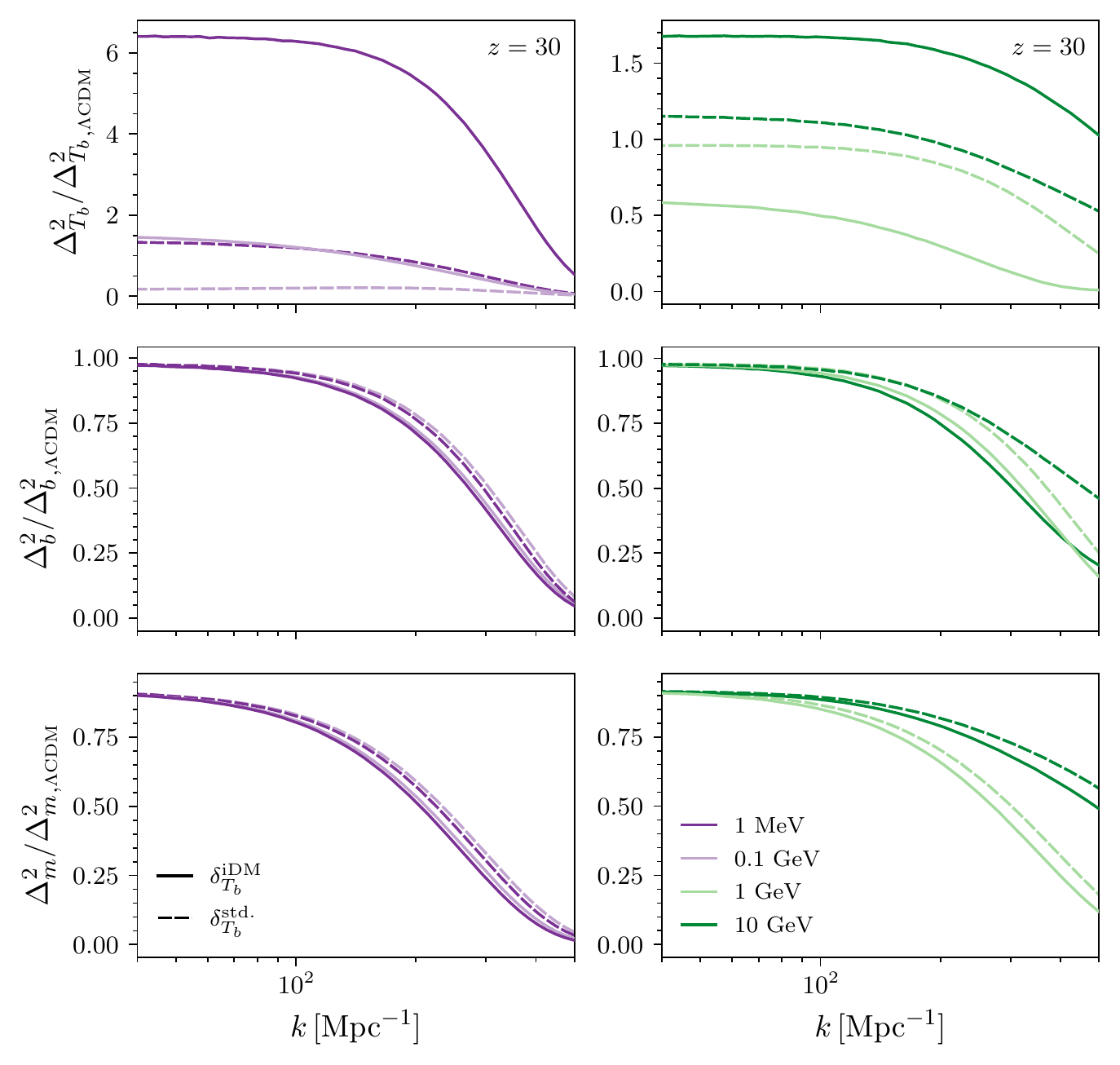}
\centering{\caption{\label{fig:power_spectra_SS} Ratio (with respect to $\Lambda$CDM) of the small-scale power spectra of baryon temperature (top), baryon density (middle) and matter density (bottom) fluctuations at $z=30$ for scales $k= 40$--$500 \, \rm Mpc^{-1}$. We compare the result when including (solid) and neglecting (dashed) the novel contributions to perturbed recombination from DM-baryon interactions for each DM mass, finding additional suppression of small-scale baryon and matter density perturbations in the former case.
}}
\end{figure}

In addition to the almost scale-independent enhancement of the temperature fluctuations, there are several effects that alter the power spectra only at scales $k\gtrsim 10$ Mpc$^{-1}$. First, there is the averaged effect due to the relative bulk velocities; this effect is qualitatively similar to $\Lambda$CDM cosmologies, featuring a suppression of fluctuations around the baryonic Jeans scale. However, DM-baryon interactions lead to a further suppression of power and a series of damped oscillations at very small scales. We find that averaging over relative bulk velocities can modify the shape of this oscillatory feature.

Second, the changes to the temperature fluctuations sourced by DM-baryon interactions are imprinted on the small-scale density fluctuations through the sound speed terms. The dark matter fluid sound speed term now has an additional contribution due to the presence of $\delta_{T_\chi}$, which acts to suppress the DM density perturbations at small scales for all cases since $\delta_{T_\chi}$ is always positive. Similarly, $\delta_{T_b}$ has new contributions, which modify the baryon fluid sound speed term: the cooler (hotter) baryon temperature in overdensities for sub-GeV masses ($m_\chi \gtrsim 1$ GeV)  tends to enhance (suppress) the baryon density power spectrum. However, due to the coupling of the two fluids, the suppressing effect on the DM due to $\delta_{T_\chi}$ dominates in all cases; we find that the baryon power spectrum is increasingly suppressed due to the contributions to perturbed recombination from DM-baryon scattering for all masses considered. The effect is smaller for sub-GeV masses because of the opposite trends in the effects of $\delta_{T_b}$ and $\delta_{T_\chi}$ described above. Of course, the new contributions to the sound speed terms result in a larger suppression of the small-scale total matter power spectrum, too. Consequently, the temperature power spectrum is also affected at small scales through its dependence on the density perturbations.

We illustrate this effect in figure~\ref{fig:power_spectra_SS}, showing the ratio of the small-scale baryon temperature, baryon density and total matter density power spectra with respect to $\Lambda$CDM at $z=30$, comparing the predictions for DM-baryon scattering cosmologies with (solid) and without (dashed) including the new contributions to perturbed recombination. For $k\lesssim 100$ Mpc$^{-1}$ we find a similar scale-independent enhancement of the baryon temperature power spectrum as demonstrated in figure~\ref{fig:PS_Tgas_z30_z50}. However, this power spectrum drops at higher $k$ due to the DM-baryon interactions. This suppression is faster when all the terms in the temperature perturbations are included, since the contributions to the sound speed terms increase. This can be seen comparing the panels corresponding to the baryon temperature and the baryon density power spectra. The relative suppression of baryon density perturbations, including all contributions to perturbed recombination, is greater for DM masses $\gtrsim 1$ GeV due to the combined effect of increased contributions to both the DM and baryon sound speed terms (e.g., for a 10 GeV DM, there is a further $\sim 10\%$ suppression of the baryon power spectrum at $k\sim 200$ Mpc$^{-1}$, compared to a little over $\sim 5\%$ for sub-GeV masses at this scale.)

On the other hand, the DM temperature perturbations $\delta T_\chi$ are larger for lighter DM masses; this results in a greater suppression of the DM density perturbations, and therefore the total matter power spectrum, at small scales. For sub-GeV masses, we find that the total matter power spectrum is suppressed by $\sim 10$--$20\%$ over scales $k\sim 200$--$300$ Mpc$^{-1}$ relative to the case when neglecting the temperature perturbations sourced by DM-baryon scattering. For heavier DM, the small-scale suppression is less steep; over the same scales, the matter power suppression ranges from $\sim 5$--$15\%$ for $m_\chi = 1$ GeV and from $\sim \rm{few}$--$10 \%$ for $m_\chi = 10$ GeV, compared to neglecting the new contributions.

Overall, depending on the DM mass, we find that including the terms sourced by DM-baryon scattering in the perturbed recombination leads to an additional $\sim 5$--$10\%$ suppression of the total matter power spectrum relative to $\Lambda$CDM for scales $k \gtrsim 200$ Mpc$^{-1}$. These results will also depend on the strength of the interaction cross section $\sigma_0$. We discuss the effect varying $\sigma_0$ for a 1 GeV mass in appendix~\ref{sec:vary_f_and_sigma}. As one might expect, the additional small-scale suppression 
that results from our full treatment of the perturbed recombination scales inversely with the scattering cross section, and the effect is less pronounced as the interaction strength weakens. However, there is still an extra suppression at the few percent level even for a value of $\sigma_0$ one order of magnitude below current limits. 

\subsection{Large-scale enhancement of baryon temperature power spectrum}
\begin{figure}[tbp]
\centering
\includegraphics{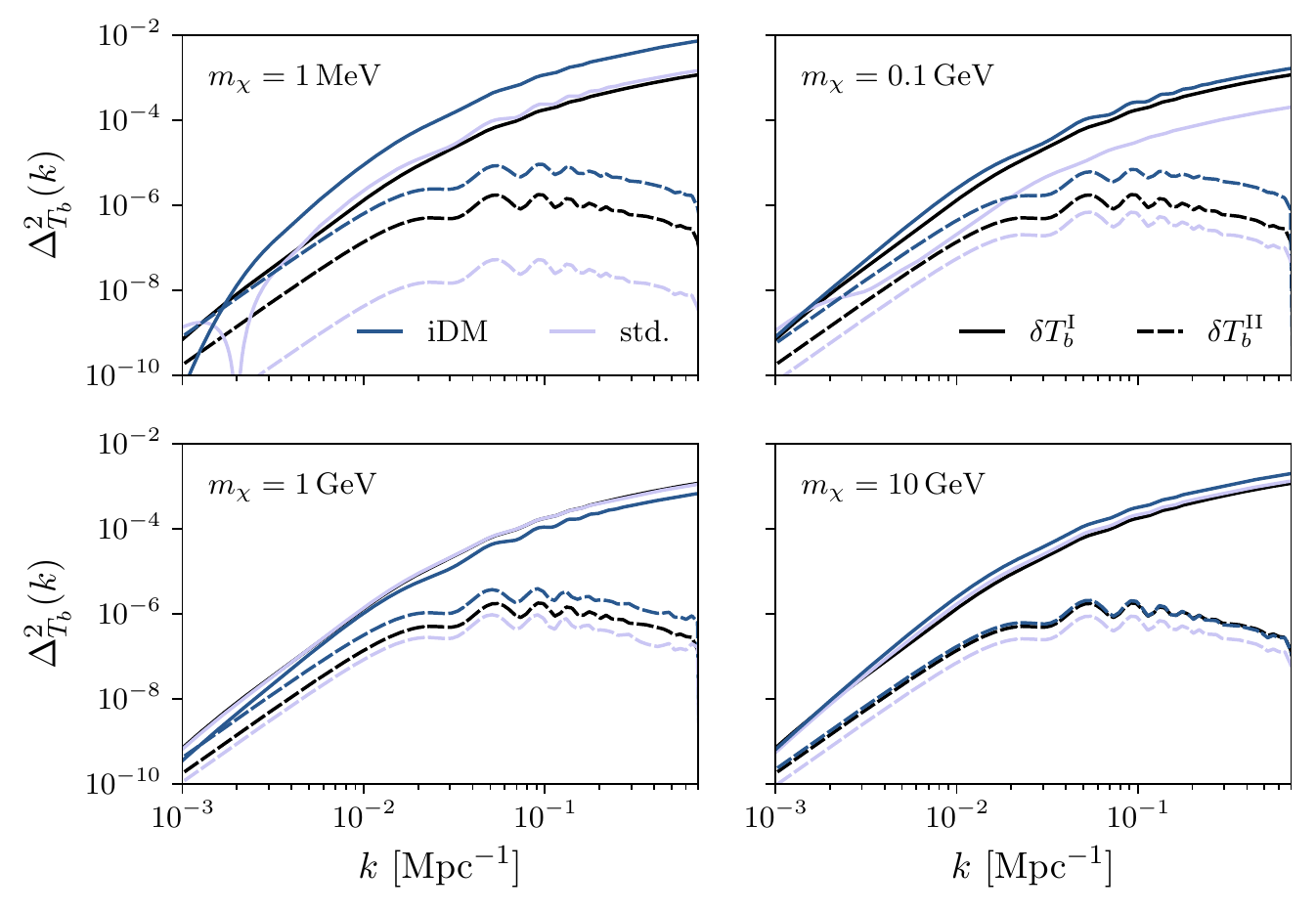}
\centering{\caption{\label{fig:Delta_TgasI_II} Quadratic correction to large-scale gas temperature fluctuations at $z=30$, including the DM-baryon scattering contribution to perturbed recombination (iDM) and neglecting them (std.). We show the $\Lambda$CDM prediction in black for reference.  Note that for $m_\chi \leq 1$ GeV, the temperature fluctuations $\delta_{T_b}$ can change sign from negative to positive at certain values of $z$; this explains the ``dip'' feature in the monopole evident for the 1 MeV case. For each DM mass we set the momentum-transfer cross section to the values found in table~\ref{tab:sigma0}.}}
\end{figure}

Finally, we study the large-scale quadratic corrections to the baryon temperature power spectrum. In the case of $\Lambda$CDM, this correction amounts to a $\sim 10\%$ enhancement of power at $k \lesssim 0.01 \, \rm{Mpc}^{-1}$ for $z=30$~\cite{Ali-Haimoud:2013hpa}. Scattering between DM and baryons can enhance the quadratic contribution to the large-scale temperature fluctuations (both in absolute terms and relative to the linear power spectrum), as we illustrate in figure~\ref{fig:Delta_TgasI_II}. For example, for a 1 GeV DM mass, the large-scale temperature fluctuations are enhanced by $\sim40\%$ at $z=30$ for $k \sim 0.01 \, \rm{Mpc}^{-1}$ and increase at larger scales (becoming comparable to the linear contribution at $k \sim 0.001\, \rm{Mpc}^{-1}$). Neglecting the temperature fluctuations sourced by DM-baryon interactions not only significantly mis-estimates the linear power spectrum, but the second order contribution at large scales as well. 

\section{Consequences of temperature perturbations in the presence of DM-baryon scattering}\label{sec:4}
In the previous sections, we have derived the new contributions to perturbed recombination arising from DM-baryon scattering and their direct effects on the evolution of cosmological perturbations, with special attention to the temperature and density power spectra. While the newly-derived contributions do not have an impact on CMB power spectra, the modified baryon temperature and growth of structure at small scales may have sizable signatures in other observables that are sensitive to the abundance of light collapsed structures and the baryonic gas. 

It is expected that the first stars form at $z\sim 30$--$40$ in the densest volumes of the Universe where gas collapses into dark matter halos and cools. As discussed above, accounting for the temperature perturbations arising from the DM-baryon scattering results in a $\sim 10\%$ suppression of the baryon and matter density power spectra with respect to the (already suppressed) previous predictions. This suppression occurs at the scales of interest for the formation of the dark matter halos that could host the first stars. Therefore, the DM-baryon interactions may delay cosmic dawn even more than originally predicted by first estimates that neglected the contributions to the temperature perturbations, which should have a very strong impact in the HI line-intensity mapping observables during cosmic dawn~\cite{Das:2017nub,Escudero:2018thh,Safarzadeh:2018hhg,Lopez-Honorez:2018ipk,Munoz:2019hjh,Munoz:2020mue,Jones:2021mrs, Giri:2022nxq, Sarkar:2022dvl}. Moreover, the modified temperature perturbations may induce additional spatial correlations in the clustering of the first stars. At the very least, our results show that the inclusion of all relevant sources of temperature perturbations is very likely to modify the initial conditions at the onset of cosmic dawn. Nonetheless, it will arguably still play an important role once sources of radiation switch on, modifying the predicted evolution of the cosmic dawn and the epoch of reionization. Given the complexity of such processes and the expected highly non-trivial interplay between the effects of the DM-baryon scattering (both on densities and temperature) and astrophysical processes, we leave a dedicated study of this to future work. In any case, we anticipate that these contributions will increase the sensitivity of experiments targeting both the high-redshift HI power spectrum -- such as HERA~\cite{DeBoer:2016tnn} and SKA~\cite{SKA:2018ckk} --  and the sky-averaged measurements -- e.g. EDGES~\cite{Bowman:2018yin}, SARAS~\cite{SARAS} and LEDA~\cite{LEDA} -- to DM-baryon scattering models.

As an example, we study the effect on the HI line-intensity mapping from the dark ages, before the formation of the first stars, which can be computed from the evolution equations in section~\ref{sec:boltz}. We first briefly review the computation of the HI line-intensity mapping observables from the dark ages, drawing heavily from refs.~\cite{Ali-Haimoud:2013hpa, Lewis:2007kz}; we refer the interested reader to these works for further details. Then, we show the predicted power spectra. We note that for $z\lesssim50$, non-linearities become important in the HI fluctuations and can affect the small-scale power spectrum at the many percent level~\cite{Lewis:2007kz}. However, we emphasise that the aim of this work is to demonstrate the importance of including DM-baryon interactions in the temperature perturbations (not just the background) when evaluating the HI signal for these models. For this purpose it is sufficient to limit ourselves to a linear treatment, but non-linear effects should be included for a high-precision prediction of the signal and comparison with data.

\subsection{HI line intensity and fluctuations} \label{sec:21cm}
The spin temperature $T_s$ of the baryon gas is defined through the ratio of the populations of neutral hydrogen in triplet to singlet state, and can be computed, in the limit $T_\star \ll T_b, T_\gamma$, as
\begin{equation}\label{eq:Ts}
T_s = T_\gamma + (T_b-T_\gamma)\frac{C_{10}}{C_{10}+A_{10}\frac{T_b}{T_\star}},
\end{equation}
where $C_{01}$ and $C_{10}$ are the upward and downward collision transition rates, respectively~\cite{Lewis:2007kz}, $A_{10} \approx 2.85 \times 10^{-15} \, \rm s^{-1}$ is the spontaneous Einstein decay rate of the HI transition, and $T_\star \approx 0.068 \rm K $ is the energy difference between the two spin states. 

The HI brightness temperature $T_{\rm HI}$, defined as the difference between the brightness of the HI radiation field and the background CMB radiation field, is equivalent to the line intensity. The observed redshifted brightness temperature today of the HI radiation emitted at redshift $z$ is given by
\begin{equation}\label{eq:T21}
T_{\rm HI}=\frac{(T_s-T_\gamma)(1-e^{-\tau})}{1+z} \approx \frac{(T_s-T_\gamma)}{1+z}\tau,
\end{equation}
where we have used the optically thin limit, valid during all of the dark ages, and $\tau$ is the  optical depth for the HI radiation.

We compute the HI brightness temperature fluctuations to second order in the hydrogen density, temperature perturbations, and ionization fraction, neglecting fluctuations in the photon temperature. Accounting for redshift-space distortions, the observed brightness temperature fluctuations to second order are 
\begin{equation}\label{eq:coeffs_T21}
\begin{split}
 \delta T_{\rm HI}^{\rm obs} ={}& -\overline{T}_{\rm HI}\delta_v + \mathcal{T}_H\delta_b+\mathcal{T}_T\delta_{T_b}+\mathcal{T}_x\delta_{x_e}  \\
&\quad+ \mathcal{T}_{HH}\delta_b^2 + \mathcal{T}_{TT}\delta_{T_b}^2 + \mathcal{T}_{xx}\delta_{x_e}^2+\mathcal{T}_{HT}\delta_b\delta_{T_b}+\mathcal{T}_{Hx}\delta_{x_e}\delta_b+\mathcal{T}_{Tx}\delta_{T_b}\delta_{x_e},
\end{split}
\end{equation}
where $\overline{T}_{\rm HI}$ is the average brightness temperature, $\delta_v\equiv\partial_\parallel v_\parallel/H$ is a dimensionless small quantity (where $\partial_\parallel v_\parallel$ is the line-of-sight gradient in proper space of the component of the peculiar velocity along the line of sight), and we assume $\delta_H \approx \delta_b$. The coefficients $\mathcal{T}_i$ are functions of redshift only (for explicit expressions see ref.~\cite{Pillepich:2006fj}).

The monopole source can be defined as 
\begin{equation}
\delta_s\equiv \frac{\mathcal{T}_H\delta_b+\mathcal{T}_T\delta_{T_b}^{\rm{I}}}{\overline{T}_{\rm HI}}
\end{equation}
and the total contribution of quadratic terms (recalling that $\delta_{T_{b}} = \delta^{\rm{I}}_{T_{b}} + \delta^{\rm{II}}_{T_{b}}$ contains quadratic terms itself) is defined as 
\begin{equation}\label{eq:quad_corr_dT21}
\delta T_{\rm HI}^{\rm{II}}\equiv\mathcal{T}_{HH}\Delta\delta_b^2+\mathcal{T}_{TT}\Delta\delta_{T_b}^2+\mathcal{T}_{HT}\Delta\delta_b\delta_{T_b}+\mathcal{T}_T\delta_{T_b}^{\rm{II}}\,,
\end{equation}
where we have neglected terms involving the ionization fraction perturbations because they are much smaller,  $\Delta\delta^2$ is the fluctuation around the mean of $\delta^2$, and relativistic terms are neglected. Therefore, the perturbations of the observed brightness temperature are
\begin{equation}
\delta T_{\rm HI}^{\rm{obs}} = \overline{T}_{\rm HI}(\delta_s - \delta_v)+\delta T_{\rm HI}^{\rm{II}} 
= \overline{T}_{\rm HI}(\delta_s - \frac{\mu^2\theta_b}{H})+\delta T_{\rm HI}^{\rm{II}}\, , 
\end{equation}
where $\mu = \bm{\hat{n}} \cdot \bm{\hat{k}}$. At small scales we only consider the first term, while at large scales the quadratic corrections are uncorrelated with the linear terms. 

\begin{figure}[tbp]
\centering
\includegraphics{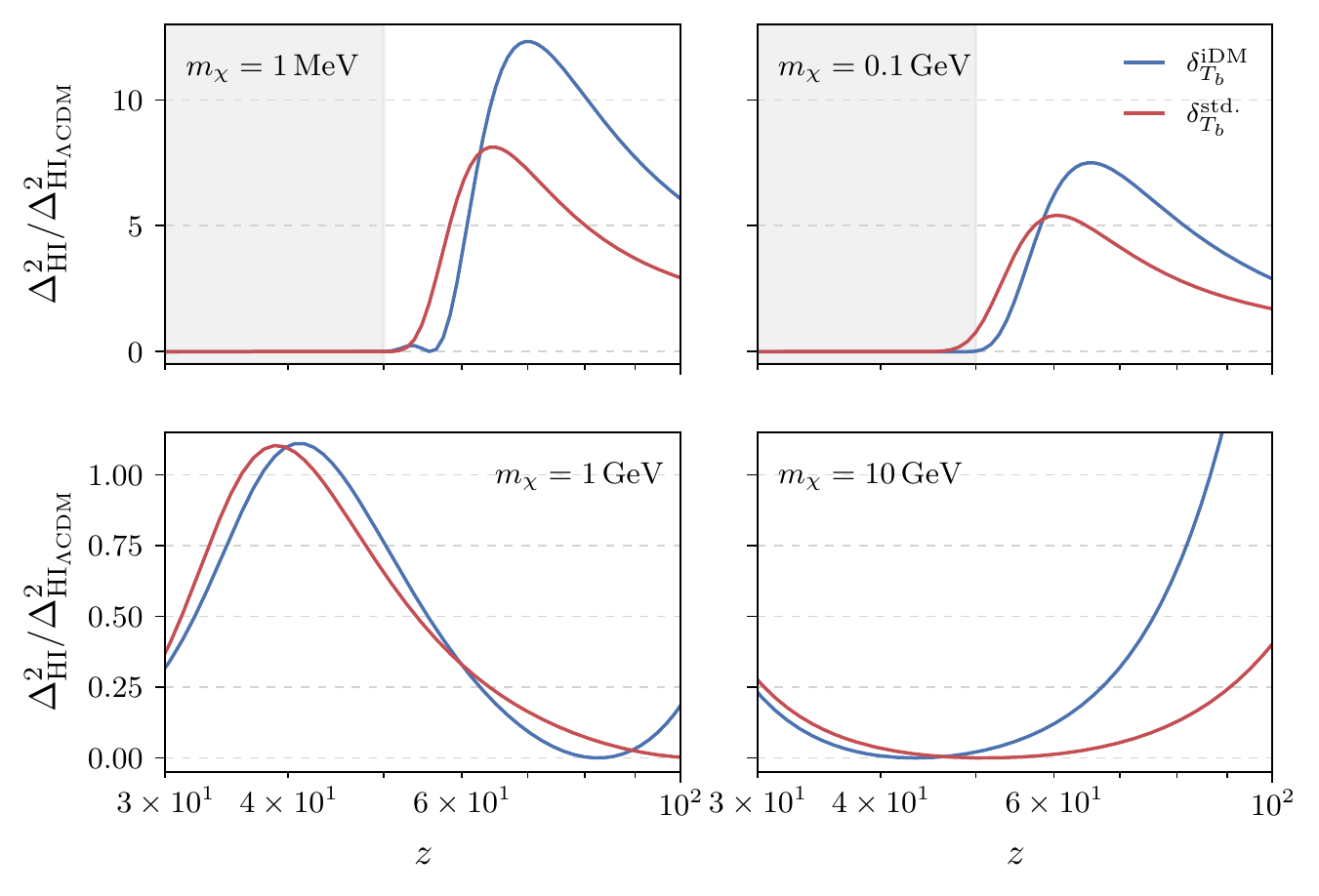}
\centering{\caption{\label{fig:21PS_monop_res} Ratio (with respect to $\Lambda$CDM) of the HI power spectrum as a function of redshift at $ k = 1\, \rm{Mpc}^{-1}$. We show the result when including the additional temperature fluctuations due to DM-baryon interactions (blue, iDM) compared to neglecting them (red, std.) For each DM mass, the value of the cross section $\sigma_0$ is found in table~\ref{tab:sigma0}. Note that for sub-GeV masses, there is effectively no HI absorption line $z \lesssim 50$, indicated by the shaded region.}}
\end{figure}

In figure~\ref{fig:21PS_monop_res}, we illustrate the effect of DM-baryon scattering on the HI power spectrum as a function of redshift for each DM mass. We show the ratio of the power spectra sourced by the monopole at $k=1$ Mpc$^{-1}$ with respect to $\Lambda$CDM, comparing the result when including (blue) and neglecting (red) the new contributions to the gas temperature fluctuations from DM-baryon scattering. The primary effect of including the additional gas temperature fluctuations is to modify the overall amplitude of the HI power spectrum over a wide range of scales and redshifts. This is not surprising, since this was also the main effect at large scales for the baryon gas temperature power spectrum which is one of the leading contributions to $\delta T_{\rm HI}$ at linear order. 

For $m_\chi < 1$ GeV we find a large enhancement at high redshifts, peaking at $z\sim 70$, compared to neglecting the effects of DM-baryon interactions in the perturbed recombination. Note that for sub-GeV masses, the cooling of the baryon gas temperature drives the mean brightness temperature $\overline{T}_{\rm HI} \rightarrow 0$ much faster, and there is effectively no HI absorption line below $z \sim 50$. For a 1 GeV DM mass, the amplitude of the HI power spectrum is enhanced for $z \gtrsim 40$ due to the enhanced baryon temperature fluctuations, except close to $z \sim 80 $ where there is a near cancellation of monopole terms ($\mathcal{T}_H\delta_b + \mathcal{T}_T\delta^{\rm{I}}_T \sim 0$); below $z\sim40$, the HI power spectrum is suppressed due to the suppression of the baryon temperature fluctuations. Finally, we see that for a 10 GeV DM the HI power spectrum is enhanced $z \gtrsim 50$; below this redshift, even though the baryon temperature fluctuations are enhanced, this results in a closer cancellation of the monopole terms and the amplitude of the HI power spectrum is actually suppressed (for the benchmark $\sigma_0$).

\begin{figure}[tbp]
\centering
\includegraphics{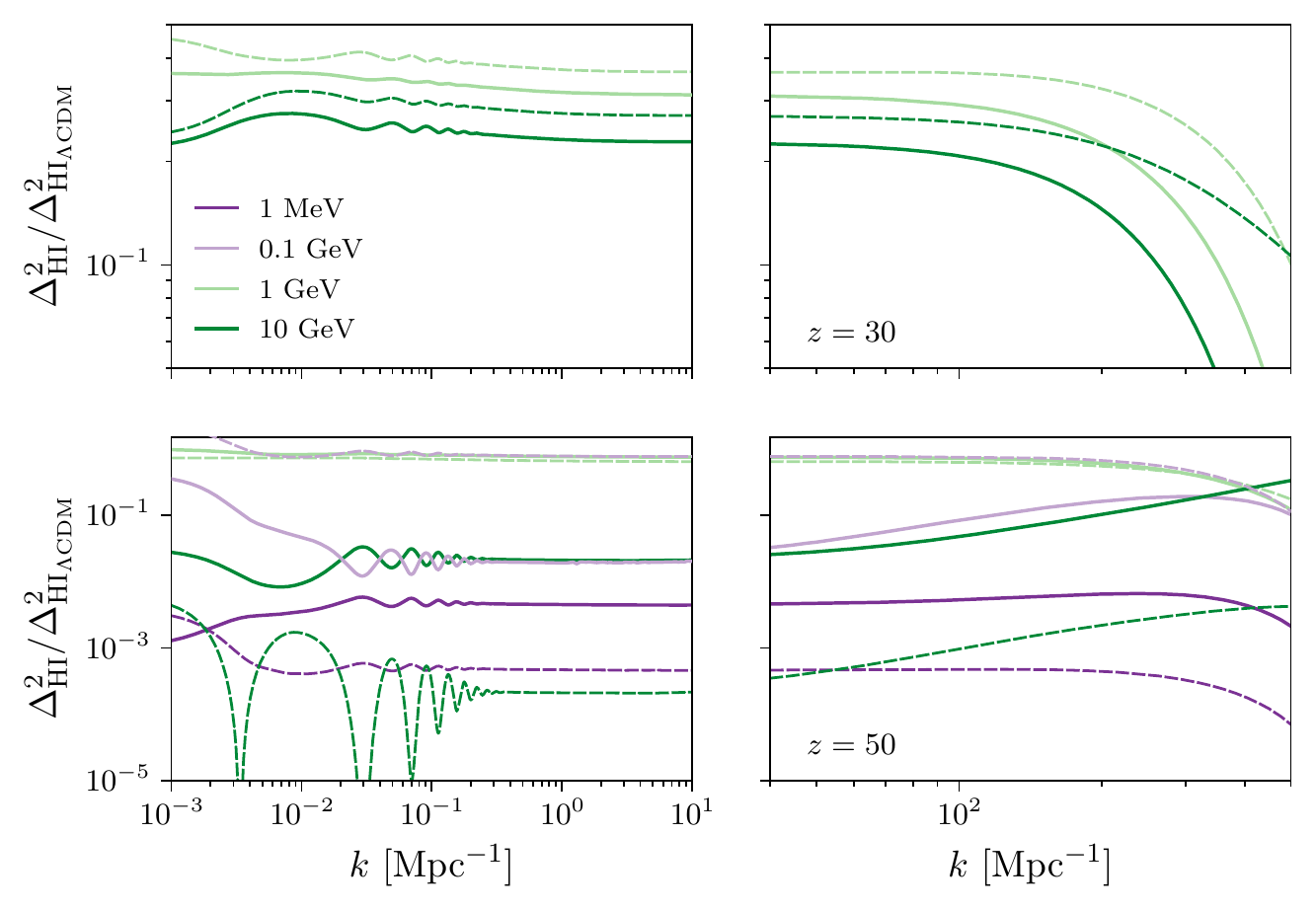}
\centering{\caption{\label{fig:P21_ratio} Ratio (with respect to $\Lambda$CDM) of the power spectrum of HI fluctuations at $z=30$ (top) and $z=50$ (bottom). On the left, we show the large-scale power spectrum both including (solid) and neglecting (dashed) the DM-baryon scattering contributions to perturbed recombination until $k= 10 \, \rm Mpc^{-1}$. On the right, we show the same for scales $k = 40$--$500 \, \rm Mpc^{-1}$.
}}
\end{figure}

In the left panels of figure~\ref{fig:P21_ratio} we show the ratio of the large-scale HI power spectrum relative to $\Lambda$CDM at $z=30$ and 50 up to $k = 10 \, \rm Mpc^{-1}$. At $z=30$, the amplitude of HI power spectrum for $ m_\chi \geq $ 1 GeV is suppressed by $\sim 15\%$ compared to the case when considering standard Compton scattering only in the temperature perturbations. Moreover, we see that the additional gas temperature fluctuations from DM-baryon interactions can enhance or modify the BAO feature in the power spectrum depending on which term in eq.~\eqref{eq:coeffs_T21} dominates. Observing this feature could help to distinguish a DM-baryon scattering signal from other exotic physics which might modify the amplitude of the HI power spectrum. 
The right panels of figure~\ref{fig:P21_ratio} show the ratio of the HI power spectrum relative to $\Lambda$CDM at $z=30$ and 50 for scales $k \geq 40 \, \rm Mpc^{-1} $. 
The effects on smaller scales are a combination of the results at larger scales (i.e., maintaining the amplitude change shown in figure~\ref{fig:21PS_monop_res}) and the suppression of temperatures and densities at scales smaller than the Jeans scale. However, the magnitude of the monopole term grows significantly in some cases due to the averaging over relative bulk velocities and the non-trivial interplay of the terms in eq.~\eqref{eq:coeffs_T21}, contrary to the density and temperature perturbations. 

\begin{figure}[tbp]
\centering
\includegraphics{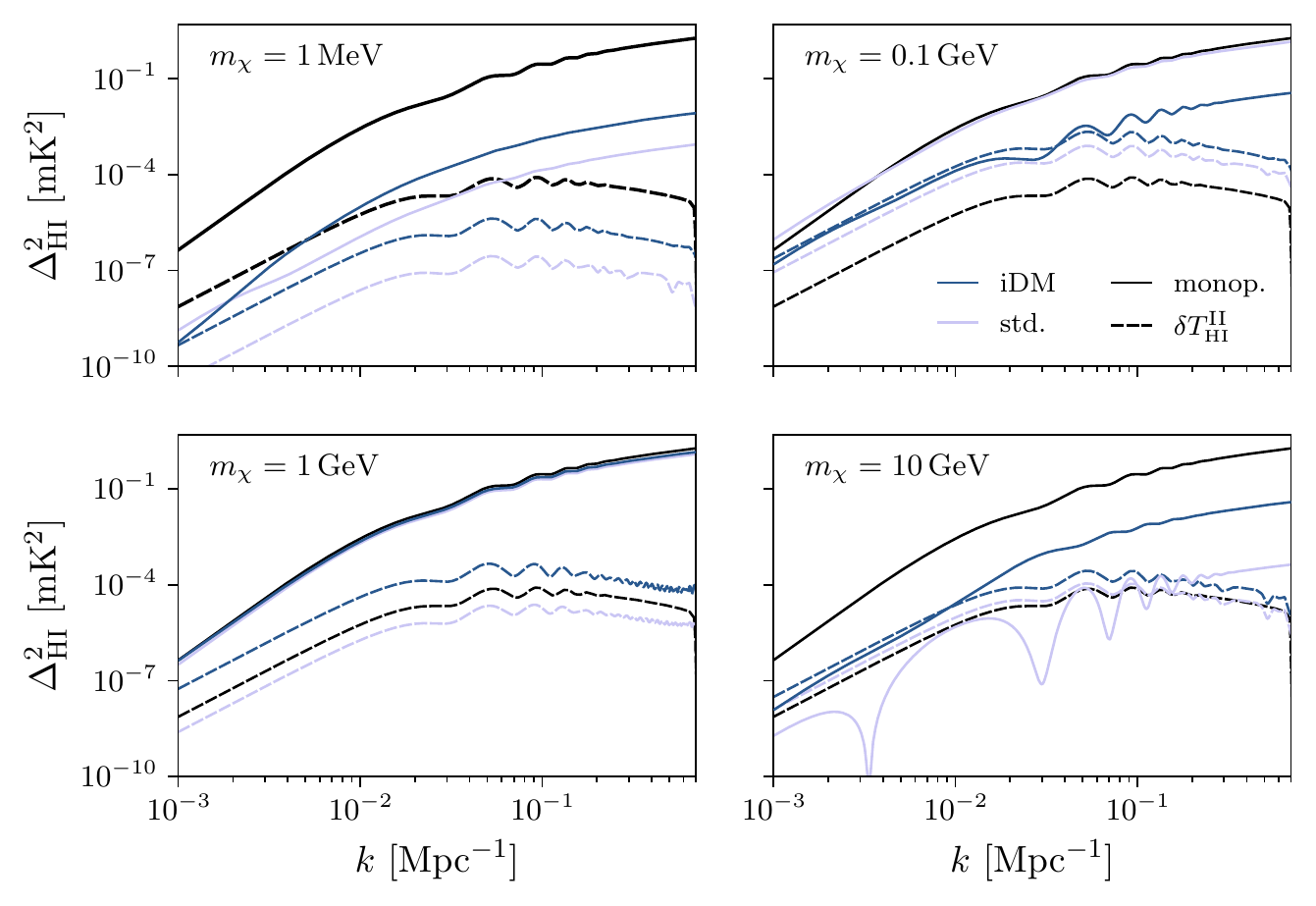}
\centering{\caption{\label{fig:Delta_T21_I_II_z50} Power spectrum of HI fluctuations at $z=50$ comparing the standard monopole term (solid) and the quadratic contribution (dashed). We compare our full calculation (iDM), to the result when ignoring the fluctuations sourced by DM-baryon scattering (std.), and to $\Lambda$CDM (black). The interaction cross section $\sigma_0$ for each DM mass is set to the values found in table~\ref{tab:sigma0}.}}
\end{figure}

Finally, in figure~\ref{fig:Delta_T21_I_II_z50}, we show the power spectrum of the quadratic terms $\delta T^{\rm II}_{\rm HI}$ compared to the monopole fluctuations for both $\Lambda$CDM and DM-baryon scattering cases
at $z=50$. In $\Lambda$CDM, the relative contribution of quadratic terms is maximal around $z\sim 30$, contributing greater than $\sim 10\%$ of the monopole fluctuations for $k\lesssim 0.01$ Mpc$^{-1}$. 
At this redshift, the quadratic contribution is unimportant for sub-GeV masses as the monopole term is negligible, and for masses 1--10 GeV the relative contribution of quadratic terms is similar to or less than for $\Lambda$CDM. However, when there are interactions between DM and baryons, we find considerable modifications to the absolute and relative amplitudes of the quadratic contributions. In particular, the relative contribution of the quadratic corrections can grow significantly at higher redshifts. The different redshift dependence of the relative contribution of these corrections to the HI and baryon temperature power spectra is due to the fact that the HI signal depends non-trivially on other quantities too (see eq.~\eqref{eq:quad_corr_dT21}). 

For both 0.1 and 10 GeV masses, the relative contribution peaks around $z\sim 50$, becoming comparable to if not larger than the signal sourced by the monopole for scales $k\lesssim 0.01$ Mpc$^{-1}$. A similar behaviour is found for $m_\chi = 1$ MeV but at slightly higher redshift (not shown). In all cases, neglecting the contributions from DM-baryon interactions in the temperature fluctuations significantly mis-estimates the amplitude of all contributions to the HI power spectrum. 

\section{Conclusions}\label{sec:conc}

We have considered the cosmological effects of dark matter that features velocity-dependent elastic interaction with protons, where the momentum-transfer cross section is parameterised as a power law of the relative particle velocity, $\sigma(v) = \sigma_0 v^{n}$. In particular, we focused on understanding the effects of the perturbed baryon and dark matter temperatures, deriving for the first time the contributions arising directly from the DM-baryon collision term and studying their effects on the evolution of cosmological perturbations. We presented numerical results for Coulomb-like scattering with $n=-4$. Such interactions have been previously invoked to explain the anomalous HI signal reported by the EDGES collaboration, but are also well-motivated from a theoretical standpoint within millicharged models for dark matter.

While the effect of DM-baryon scattering on the background DM and baryon temperatures has been well-studied in the literature, this is the first time the impact at the level of the perturbations has been considered. We derived the modified perturbed recombination equations for the temperatures and the ionization fraction and evolved them self-consistently with the linear Boltzmann equations appropriate for DM-baryon scattering cosmologies until the end of the dark ages ($z\sim 30$). By doing so, we also computed how the modified DM and baryon temperature perturbations affect the growth of structure by properly accounting for their contribution to the sound speed terms in the momentum evolution of each fluid respectively. 

We found that the baryon temperature (and ionization fraction) perturbations are enhanced by up to 1--2 orders of magnitude when including the novel DM-baryon scattering contributions to perturbed recombination, for scattering cross sections within current CMB bounds. These changes are significant over a wide redshift throughout the dark ages, and largely scale-independent for $k \lesssim 100 \, \rm Mpc^{-1}$. Moreover, the DM-baryon interactions result in non-negligible DM temperature perturbations as well. These additional contributions to the DM and baryon temperature perturbations also affect the growth of density perturbations at the Jeans scale through the sound speed terms in the Boltzmann equations; we found a further $\sim 5\%$--$10\%$ suppression of the baryon and matter power spectra at $k\sim 200\, {\rm Mpc^{-1}}$ relative to $\Lambda$CDM at $z=30$, compared to the prediction without including these effects (depending on the DM mass and scattering cross section). 

The effects of DM-baryon scattering on the temperature perturbations, and the resulting suppression of small-scale structures, have important implications for the modelling and interpretation of cosmological  
observables sensitive to the gas temperature and the abundance of light collapsed objects. The evolution of cosmic dawn and reionization is highly sensitive to the early-time star formation history, which in turn depends upon the fraction of gas collapsed into halos where it can sufficiently cool and form stars. Therefore, our results will be critical to accurately predict the initial conditions of cosmic dawn and we anticipate this to have a significant impact on the reionization history, delaying the milestone epochs of cosmic dawn, X-ray heating and reionization. This will provide future observations with additional sensitivity to DM-baryon scattering models. One example of such an observable is HI intensity mapping. To illustrate this point, we computed the HI power spectrum from the dark ages including the newly derived contributions to the temperature perturbations from DM-baryon scattering. 
We found that the amplitude of the HI power spectrum is modified by tens of percent at $z=30$, and up to factor of a few at higher redshifts. The effect of the interactions is also expected to alter the global HI signal~\cite{driskell_in_prep}.

We have considered the end of the cosmic dark ages to be at $z\sim 30$. However, if cosmic dawn begins earlier, our main conclusions still hold; the novel effects of DM-baryon scattering on the temperature perturbations and the matter power spectra are significant at higher redshift, which is shown explicitly for $z=50$ in Appendix~\ref{sec:z50}. We emphasise that the (previously neglected) temperature perturbations sourced by DM-baryon scattering derived in this work are required to properly model the initial conditions for cosmic dawn. Once the first stars form, their radiation will become the dominant heating source over the Compton or DM-baryon scattering contributions, and a dedicated study of the interplay of these different effects is necessary to predict the evolution of the cosmic dawn signal. 
Our results strongly motivate further investigation into how these effects would impact the cosmic dawn signal beyond $z=30$, which are likely to further improve the constraining power of forthcoming probes of the high-redshift Universe to DM-baryon interactions.

\acknowledgments
KS acknowledges support from the European Union’s Horizon 2020 research and innovation programme under
the Marie Sk\l{}odowska-Curie grant agreement No.~713673. KS is also supported by an INPhINIT fellowship from "la
Caixa" Foundation (ID 100010434), grant code LCF/BQ/DI17/11620047. 
JLB is supported by the Allan C. and Dorothy H. Davis Fellowship. 
KB acknowledges support from the National Science Foundation (NSF) under Grant No.~PHY-2112884.
VG acknowledges support from the NSF under Grant No. PHY-2013951.
LV acknowledges support from the European Union's Horizon 2020 research and innovation programme ERC (BePreSysE, grant agreement 725327). 
Funding for this work was partially provided by project PGC2018-098866-B-I00 MCIN/AEI/ 10.13039/501100011033 y FEDER “Una manera de hacer Europa”, and the “Center of Excellence Maria de Maeztu 2020-2023” award to the ICCUB (CEX2019-000918-M funded by MCIN/AEI/ 10.13039/501100011033).

\appendix
\counterwithin{figure}{section}
\section{Results at $z=50$}\label{sec:z50}
\begin{figure}[tbp]
\centering
\includegraphics{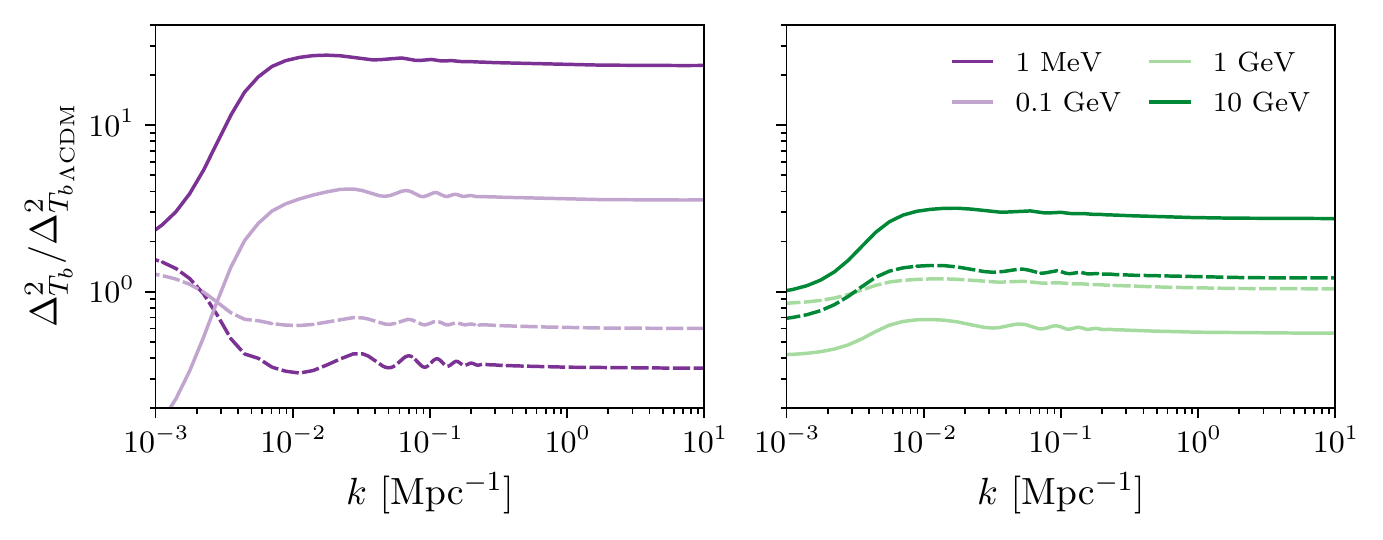}
\centering{\caption{\label{fig:PS_Tgas_z50} Ratio of the linear power spectra (with respect to $\Lambda$CDM) of the gas temperature fluctuations at $z=50$ for several DM masses, including (solid) and neglecting (dashed) fluctuations sourced by DM-baryon interactions, shown until $k = 10\, \rm Mpc^{-1}$.}}
\end{figure}

\begin{figure}[tbp]
\centering
\includegraphics{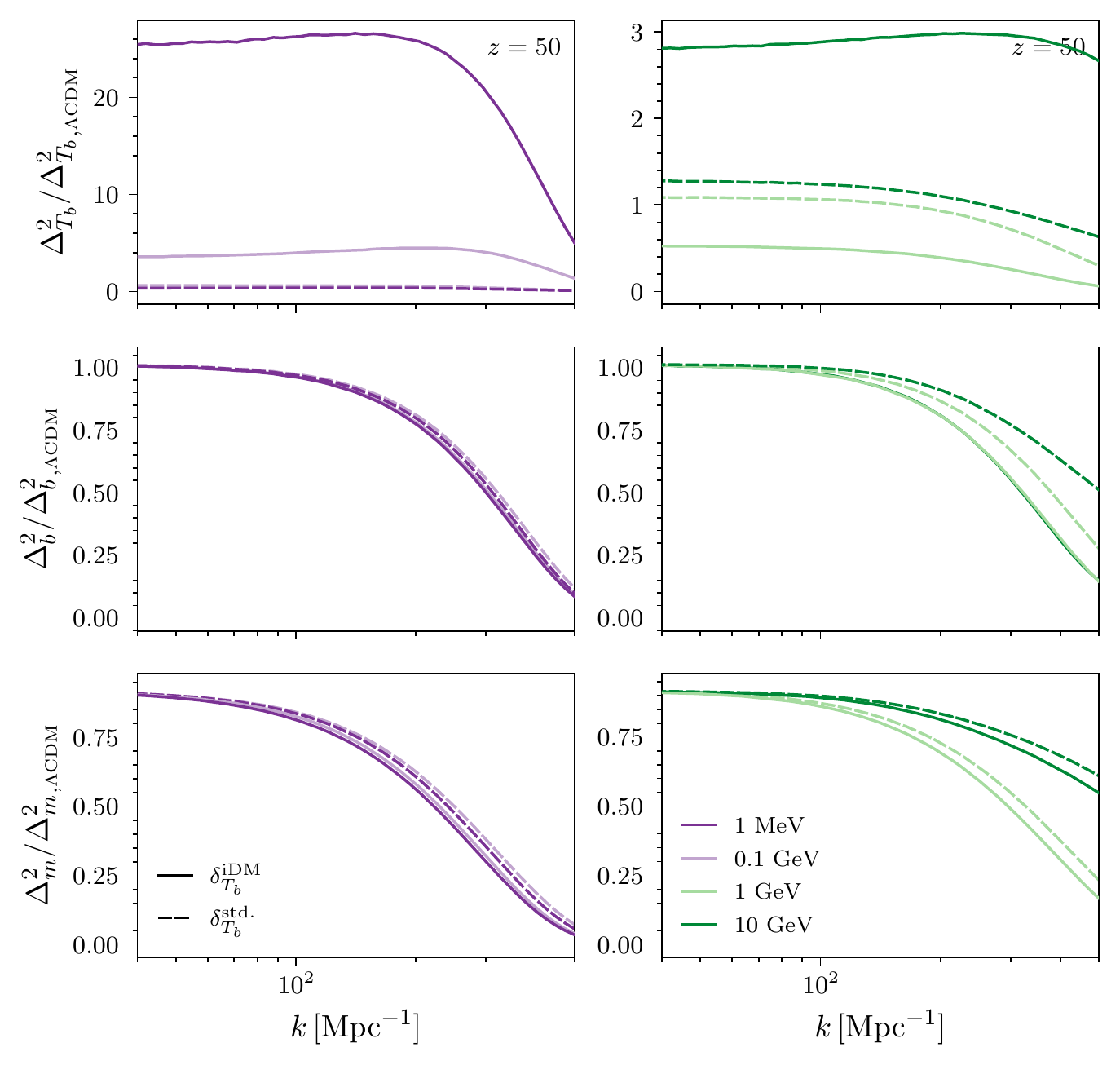}
\centering{\caption{\label{fig:power_spectra_SS_z50} Ratio (with respect to $\Lambda$CDM) of small-scale power spectra of baryon temperature (top), baryon density (middle) and matter density (bottom) fluctuations at $z=50$ for scales $k= 40$--$500 \, \rm Mpc^{-1}$. We compare the result when including (solid) and neglecting (dashed) additional contributions to perturbed recombination from DM-baryon interactions for each DM mass.}}
\end{figure}

In the main text we show results at $z=30$ which we consider to mark the end of the dark ages and the beginning of cosmic dawn. However, as demonstrated in figure~\ref{fig:dTgas_xe_k1}, there are significant changes to the temperature and ionization fraction perturbations during much of the redshift range of the dark ages.
We highlight here that there are important effects beyond $z>30$ which, given the astrophysical uncertainties surrounding early stellar formation, may be relevant for modelling the initial conditions of cosmic dawn at higher redshift. Figures~\ref{fig:PS_Tgas_z50} and \ref{fig:power_spectra_SS_z50} are analogous to figures~\ref{fig:PS_Tgas_z30_z50} and \ref{fig:power_spectra_SS}, but shown for $z=50$.

\section{Varying scattering cross section and the fraction of interacting DM}\label{sec:vary_f_and_sigma}

\begin{figure}[tbp]
\centering
\includegraphics{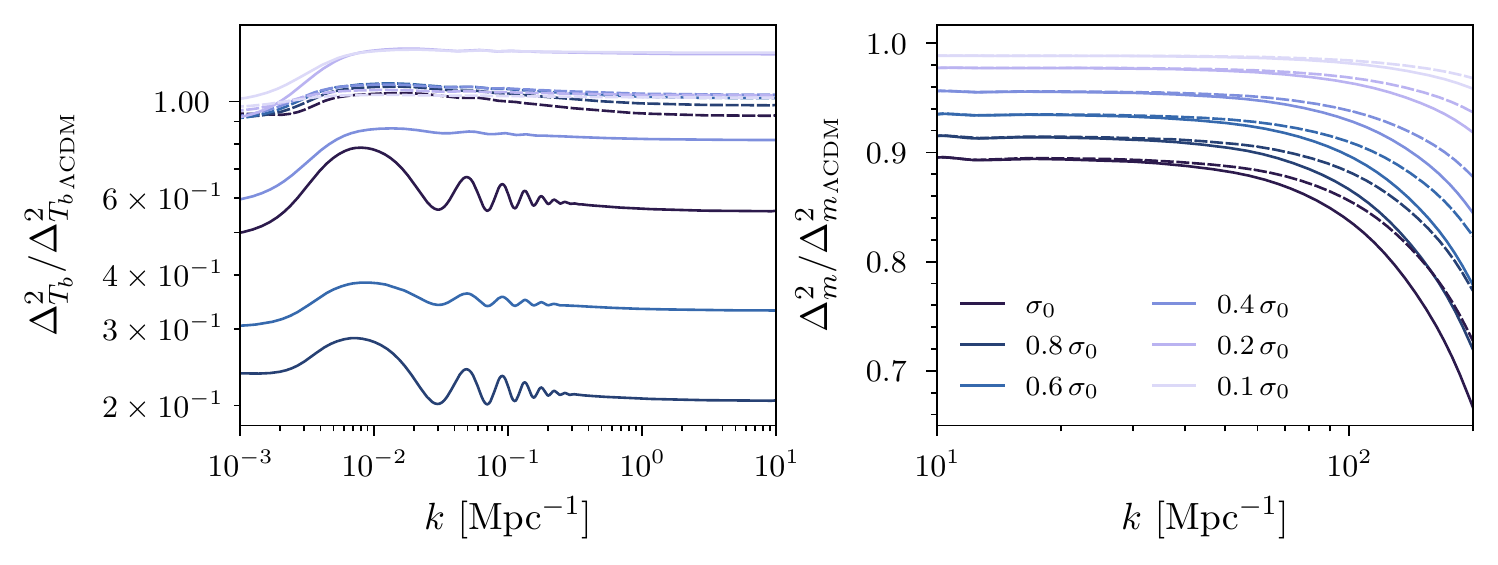}
\centering{\caption{\label{fig:1GeV_sigma} Effect of varying the scattering cross section for a DM mass of 1 GeV on the large-scale baryon temperature power spectrum (left) and the small-scale total matter power spectrum (right) at $z=30$. We compare the result when including (solid) and neglecting (dashed) the additional contributions to perturbed recombination resulting from the DM-baryon interactions for several values of $\sigma_0$ below the 95\% C.L. upper limit given in table~\ref{tab:sigma0}.}}
\end{figure}

Throughout the main text we show results for the case where all of the dark matter interacts with baryons and adopt a benchmark value of $\sigma_0$ for each DM mass, consistent with the 95\% C.L. upper limits derived in refs.~\cite{Boddy:2018wzy, Xu:2018efh, Slatyer:2018aqg}. In this appendix we vary these parameters, considering different values of $\sigma_0$ and small fractions of strongly interacting DM. 

In figure~\ref{fig:1GeV_sigma} we show how results change when reducing the strength of the interaction cross section by a numerical factor below the current upper limit for $m_\chi = 1$ GeV. We find that as we reduce $\sigma_0$, there is an intermediate range where the change in the amplitude of the temperature power spectrum at $z=30$, with respect to neglecting temperature perturbations sourced by DM-baryon scattering, actually increases significantly before decreasing again. This is due to how the evolution of the temperature perturbations $\delta_{T_b}$ depends on the different contributions in eq.~\eqref{eq:Tgas_1stO}; varying the cross section alters the balance of these competing terms and can result in either more or less enhancement or suppression of the temperature perturbations, depending on the redshift. For example, for an interaction cross section set at 80\% of our benchmark $\sigma_0$, we find that the temperature power spectrum is suppressed by almost an order of magnitude at $z=30$. This also has an impact in the large-scale HI power spectrum; the amplitude of the HI signal becomes suppressed by $\sim 25\%$ at $z=30$ relative to neglecting the DM-baryon scattering effects on the temperature perturbations (compared to $\sim 15\%$ for our benchmark $\sigma_0$). 

On the other hand, the amount of extra suppression in the total matter power spectrum due to the DM-baryon interaction contribution to the perturbations scales inversely with $\sigma_0$. For a cross section one order of magnitude weaker than current limits, there is an additional suppression of a $\sim$few \% in the total matter power spectrum at $k=200 \, \rm Mpc^{-1}$. 

\begin{figure}[tbp]
\centering
\includegraphics{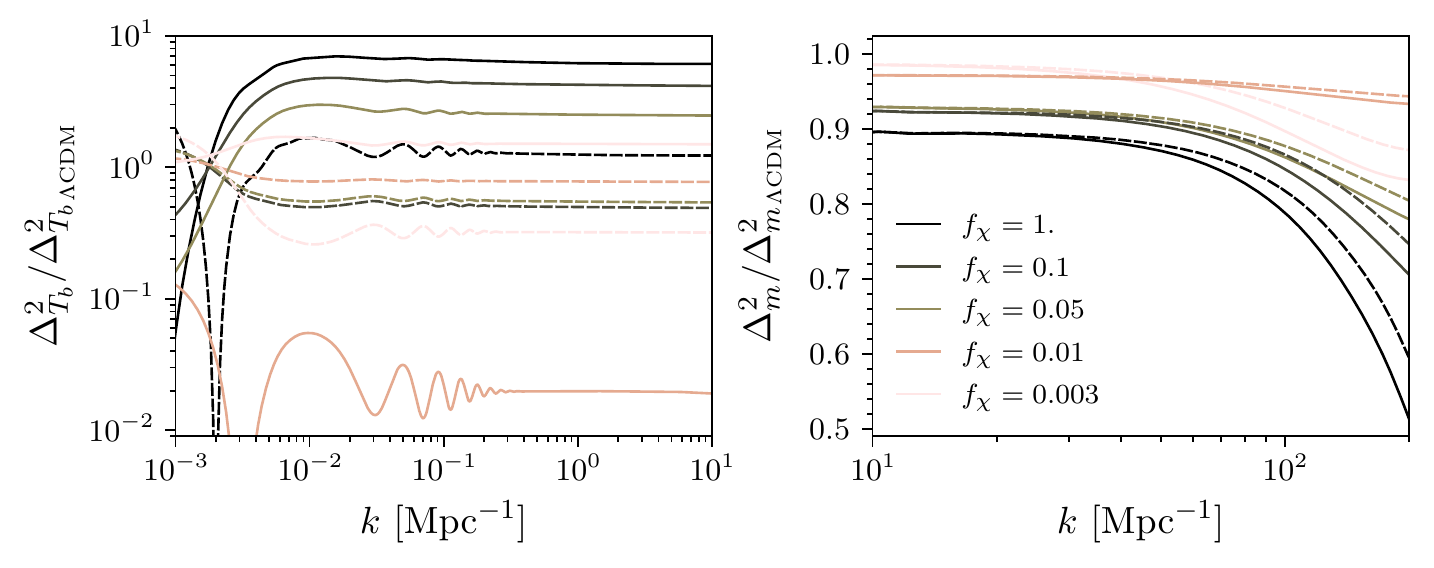}
\centering{\caption{\label{fig:1MeV_frac} 
Effect of varying the interacting fraction of DM $f_\chi$ for a DM mass of 1 MeV, with $\sigma_0$ set to the respective 95\% C.L. upper limits derived in ref.~\cite{Boddy:2018wzy} (see table~\ref{tab:sigma0_vary_f}). We show the large-scale baryon temperature power spectrum (left) and the small-scale total matter power spectrum (right) at $z=30$, comparing the result when including (solid) and neglecting (dashed) the additional contributions to perturbed recombination resulting from the DM-baryon interactions.}}
\end{figure}

Lastly, we also consider how our results change when only a fraction $f_\chi$ of the total DM density is interacting; in this regime larger values of the cross section are allowed for small interacting fractions. In figure~\ref{fig:1MeV_frac} we show results for various fractions with $\sigma_0$ set to the corresponding 95\% C.L. upper limit derived in ref.~\cite{Boddy:2018wzy} for a mass of 1 MeV (reported in table~\ref{tab:sigma0_vary_f}). 

For $f_\chi \gtrsim 2\%$ the limit on $\sigma_0$ roughly scales with $f_\chi$. In this regime, we find that the amount of extra suppression in the matter power spectrum declines with smaller $f_\chi$. On the other hand, the amplification of the temperature perturbations can be substantially increased; e.g., for $f_\chi = 0.1$, the large-scale temperature power spectrum is enhanced by almost an order of magnitude at $z=30$ when fully accounting for all contributions to perturbed recombination. The difference in scale of the impact on the baryon temperature and the matter power spectra can be explained as follows. As shown in figure~\ref{fig:dTgas_xe_k1}, each contribution to the gas temperature depends indirectly on the rest and the evolution with redshift is much richer than when only Compton heating is considered. Therefore, even if only a small part of the DM interacts, the change can be very large. However, the effect on the total matter density decreases significantly because the temperature of the (interacting) dark matter now only affects a small fraction of the total dark matter. 

The sensitivity of CMB constraints to DM-baryon interactions degrades significantly for very small interacting fractions (below $f_\chi \sim 0.4\%$~\cite{Boddy:2018wzy}), and large values of $\sigma_0$ are permitted for which the interacting sub-component of DM is tightly coupled to the baryons. 
In this regime, we find there can still be considerable modifications to the baryon temperature perturbations (for $f_\chi=0.3\%$, suppressed by almost a factor of $\sim100$ at $z=30$), while the BAO feature is markedly enhanced. Moreover, for such small tightly coupled fractions, the additional suppression in the small-scale matter power spectrum can increase again.

These results motivate further investigation and modelling of the effects of these contributions to the full cosmic dawn signal, especially in the context of millicharged DM models invoked to explain the EDGES measurement. 

\begin{table}[tbp]
\centering
\begin{tabular}{@{}cc@{}}
\toprule
 $f_\chi$ & $\sigma_0 \, [\rm{cm}^2]$ \\ \midrule
1            & $1.7 \times 10^{-41}$    \\
0.1            & $1.9 \times 10^{-40}$    \\
0.05          & $6.0\times 10^{-40}$   \\
0.01           & $5.5\times 10^{-39}$    \\
0.003           & $5.0\times 10^{-35}$    \\ \bottomrule
\end{tabular}
\caption{Benchmark values of the momentum-transfer cross section coefficient $\sigma_0$ for various interacting DM fractions $f_\chi$ for $m_\chi=1$ MeV, consistent with 95\% C.L. upper limits derived in ref.~\cite{Boddy:2018wzy}.}
\label{tab:sigma0_vary_f}
\end{table}

\bibliographystyle{utcaps}
\bibliography{21cmDMBaryon}

\providecommand{\href}[2]{#2}\begingroup\raggedright\begin{thebibliography}{10}

\bibitem{Planck:2018vyg}
{\bfseries Planck} Collaboration, N.~Aghanim {\em et~al.}, ``{Planck 2018
  results. VI. Cosmological parameters},''
  \href{http://dx.doi.org/10.1051/0004-6361/201833910}{{\em Astron. Astrophys.}
  {\bfseries 641} (2020) A6}, \href{http://arxiv.org/abs/1807.06209}{{\ttfamily
  arXiv:1807.06209 [astro-ph.CO]}}. [Erratum: Astron.Astrophys. 652, C4
  (2021)].

\bibitem{eBOSS:2020yzd}
{\bfseries eBOSS} Collaboration, S.~Alam {\em et~al.}, ``{Completed SDSS-IV
  extended Baryon Oscillation Spectroscopic Survey: Cosmological implications
  from two decades of spectroscopic surveys at the Apache Point Observatory},''
  \href{http://dx.doi.org/10.1103/PhysRevD.103.083533}{{\em Phys. Rev. D}
  {\bfseries 103} no.~8, (2021) 083533},
  \href{http://arxiv.org/abs/2007.08991}{{\ttfamily arXiv:2007.08991
  [astro-ph.CO]}}.

\bibitem{DES:2021wwk}
{\bfseries DES} Collaboration, T.~M.~C. Abbott {\em et~al.}, ``{Dark Energy
  Survey Year 3 Results: Cosmological Constraints from Galaxy Clustering and
  Weak Lensing},'' \href{http://arxiv.org/abs/2105.13549}{{\ttfamily
  arXiv:2105.13549 [astro-ph.CO]}}.

\bibitem{Heymans:2020gsg}
C.~Heymans {\em et~al.}, ``{KiDS-1000 Cosmology: Multi-probe weak gravitational
  lensing and spectroscopic galaxy clustering constraints},''
  \href{http://dx.doi.org/10.1051/0004-6361/202039063}{{\em Astron. Astrophys.}
  {\bfseries 646} (2021) A140},
  \href{http://arxiv.org/abs/2007.15632}{{\ttfamily arXiv:2007.15632
  [astro-ph.CO]}}.

\bibitem{Brout:2022vxf}
D.~Brout {\em et~al.}, ``{The Pantheon+ Analysis: Cosmological Constraints},''
  \href{http://arxiv.org/abs/2202.04077}{{\ttfamily arXiv:2202.04077
  [astro-ph.CO]}}.

\bibitem{Billard:2021uyg}
J.~Billard {\em et~al.}, ``{Direct Detection of Dark Matter -- APPEC Committee
  Report},'' \href{http://arxiv.org/abs/2104.07634}{{\ttfamily arXiv:2104.07634
  [hep-ex]}}.

\bibitem{Boveia:2018yeb}
A.~Boveia and C.~Doglioni, ``{Dark Matter Searches at Colliders},''
  \href{http://dx.doi.org/10.1146/annurev-nucl-101917-021008}{{\em Ann. Rev.
  Nucl. Part. Sci.} {\bfseries 68} (2018) 429--459},
  \href{http://arxiv.org/abs/1810.12238}{{\ttfamily arXiv:1810.12238
  [hep-ex]}}.

\bibitem{Boddy:2022knd}
K.~K. Boddy {\em et~al.}, ``{Astrophysical and Cosmological Probes of Dark
  Matter},'' in {\em {2022 Snowmass Summer Study}}.
\newblock 3, 2022.
\newblock \href{http://arxiv.org/abs/2203.06380}{{\ttfamily arXiv:2203.06380
  [hep-ph]}}.

\bibitem{Bertone_2005}
G.~Bertone, D.~Hooper, and J.~Silk, ``Particle dark matter: evidence,
  candidates and constraints,''
  \href{http://dx.doi.org/10.1016/j.physrep.2004.08.031}{{\em Physics Reports}
  {\bfseries 405} no.~5-6, (Jan, 2005) 279--390}.
  \url{http://dx.doi.org/10.1016/j.physrep.2004.08.031}.

\bibitem{Roszkowski:2017nbc}
L.~Roszkowski, E.~M. Sessolo, and S.~Trojanowski, ``{WIMP dark matter
  candidates and searches\textemdash{}current status and future prospects},''
  \href{http://dx.doi.org/10.1088/1361-6633/aab913}{{\em Rept. Prog. Phys.}
  {\bfseries 81} no.~6, (2018) 066201},
  \href{http://arxiv.org/abs/1707.06277}{{\ttfamily arXiv:1707.06277
  [hep-ph]}}.

\bibitem{Chen:2002yh}
X.-l. Chen, S.~Hannestad, and R.~J. Scherrer, ``{Cosmic microwave background
  and large scale structure limits on the interaction between dark matter and
  baryons},'' \href{http://dx.doi.org/10.1103/PhysRevD.65.123515}{{\em Phys.
  Rev. D} {\bfseries 65} (2002) 123515},
  \href{http://arxiv.org/abs/astro-ph/0202496}{{\ttfamily
  arXiv:astro-ph/0202496}}.

\bibitem{Sigurdson:2004zp}
K.~Sigurdson, M.~Doran, A.~Kurylov, R.~R. Caldwell, and M.~Kamionkowski,
  ``{Dark-matter electric and magnetic dipole moments},''
  \href{http://dx.doi.org/10.1103/PhysRevD.70.083501}{{\em Phys. Rev. D}
  {\bfseries 70} (2004) 083501},
  \href{http://arxiv.org/abs/astro-ph/0406355}{{\ttfamily
  arXiv:astro-ph/0406355}}. [Erratum: Phys.Rev.D 73, 089903 (2006)].

\bibitem{Dvorkin:2013cea}
C.~Dvorkin, K.~Blum, and M.~Kamionkowski, ``{Constraining Dark Matter-Baryon
  Scattering with Linear Cosmology},''
  \href{http://dx.doi.org/10.1103/PhysRevD.89.023519}{{\em Phys. Rev. D}
  {\bfseries 89} no.~2, (2014) 023519},
  \href{http://arxiv.org/abs/1311.2937}{{\ttfamily arXiv:1311.2937
  [astro-ph.CO]}}.

\bibitem{Gluscevic:2017ywp}
V.~Gluscevic and K.~K. Boddy, ``{Constraints on Scattering of
  keV\textendash{}TeV Dark Matter with Protons in the Early Universe},''
  \href{http://dx.doi.org/10.1103/PhysRevLett.121.081301}{{\em Phys. Rev.
  Lett.} {\bfseries 121} no.~8, (2018) 081301},
  \href{http://arxiv.org/abs/1712.07133}{{\ttfamily arXiv:1712.07133
  [astro-ph.CO]}}.

\bibitem{Boddy:2018kfv}
K.~K. Boddy and V.~Gluscevic, ``{First Cosmological Constraint on the Effective
  Theory of Dark Matter-Proton Interactions},''
  \href{http://dx.doi.org/10.1103/PhysRevD.98.083510}{{\em Phys. Rev. D}
  {\bfseries 98} no.~8, (2018) 083510},
  \href{http://arxiv.org/abs/1801.08609}{{\ttfamily arXiv:1801.08609
  [astro-ph.CO]}}.

\bibitem{Boddy:2018wzy}
K.~K. Boddy, V.~Gluscevic, V.~Poulin, E.~D. Kovetz, M.~Kamionkowski, and
  R.~Barkana, ``{Critical assessment of CMB limits on dark matter-baryon
  scattering: New treatment of the relative bulk velocity},''
  \href{http://dx.doi.org/10.1103/PhysRevD.98.123506}{{\em Phys. Rev.}
  {\bfseries D98} no.~12, (2018) 123506},
\href{http://arxiv.org/abs/1808.00001}{{\ttfamily arXiv:1808.00001
  [astro-ph.CO]}}.

\bibitem{Xu:2018efh}
W.~L. Xu, C.~Dvorkin, and A.~Chael, ``{Probing sub-GeV Dark Matter-Baryon
  Scattering with Cosmological Observables},''
  \href{http://dx.doi.org/10.1103/PhysRevD.97.103530}{{\em Phys. Rev. D}
  {\bfseries 97} no.~10, (2018) 103530},
  \href{http://arxiv.org/abs/1802.06788}{{\ttfamily arXiv:1802.06788
  [astro-ph.CO]}}.

\bibitem{Slatyer:2018aqg}
T.~R. Slatyer and C.-L. Wu, ``{Early-Universe constraints on dark matter-baryon
  scattering and their implications for a global 21 cm signal},''
  \href{http://dx.doi.org/10.1103/PhysRevD.98.023013}{{\em Phys. Rev. D}
  {\bfseries 98} no.~2, (2018) 023013},
  \href{http://arxiv.org/abs/1803.09734}{{\ttfamily arXiv:1803.09734
  [astro-ph.CO]}}.

\bibitem{Li:2018zdm}
Z.~Li, V.~Gluscevic, K.~K. Boddy, and M.~S. Madhavacheril, ``{Disentangling
  Dark Physics with Cosmic Microwave Background Experiments},''
  \href{http://dx.doi.org/10.1103/PhysRevD.98.123524}{{\em Phys. Rev. D}
  {\bfseries 98} no.~12, (2018) 123524},
  \href{http://arxiv.org/abs/1806.10165}{{\ttfamily arXiv:1806.10165
  [astro-ph.CO]}}.

\bibitem{Nadler:2019zrb}
E.~O. Nadler, V.~Gluscevic, K.~K. Boddy, and R.~H. Wechsler, ``{Constraints on
  Dark Matter Microphysics from the Milky Way Satellite Population},''
  \href{http://dx.doi.org/10.3847/2041-8213/ab1eb2}{{\em Astrophys. J. Lett.}
  {\bfseries 878} no.~2, (2019) 32},
  \href{http://arxiv.org/abs/1904.10000}{{\ttfamily arXiv:1904.10000
  [astro-ph.CO]}}. [Erratum: Astrophys.J.Lett. 897, L46 (2020), Erratum:
  Astrophys.J. 897, L46 (2020)].

\bibitem{DES:2020fxi}
{\bfseries DES} Collaboration, E.~O. Nadler {\em et~al.}, ``{Milky Way
  Satellite Census. III. Constraints on Dark Matter Properties from
  Observations of Milky Way Satellite Galaxies},''
  \href{http://dx.doi.org/10.1103/PhysRevLett.126.091101}{{\em Phys. Rev.
  Lett.} {\bfseries 126} (2021) 091101},
  \href{http://arxiv.org/abs/2008.00022}{{\ttfamily arXiv:2008.00022
  [astro-ph.CO]}}.

\bibitem{Nguyen:2021cnb}
D.~V. Nguyen, D.~Sarnaaik, K.~K. Boddy, E.~O. Nadler, and V.~Gluscevic,
  ``{Observational constraints on dark matter scattering with electrons},''
  \href{http://dx.doi.org/10.1103/PhysRevD.104.103521}{{\em Phys. Rev. D}
  {\bfseries 104} no.~10, (2021) 103521},
  \href{http://arxiv.org/abs/2107.12380}{{\ttfamily arXiv:2107.12380
  [astro-ph.CO]}}.

\bibitem{Buen-Abad:2021mvc}
M.~A. Buen-Abad, R.~Essig, D.~McKeen, and Y.-M. Zhong, ``{Cosmological
  constraints on dark matter interactions with ordinary matter},''
  \href{http://dx.doi.org/10.1016/j.physrep.2022.02.006}{{\em Phys. Rept.}
  {\bfseries 961} (2022) 1--35},
  \href{http://arxiv.org/abs/2107.12377}{{\ttfamily arXiv:2107.12377
  [astro-ph.CO]}}.

\bibitem{Maamari:2020aqz}
K.~Maamari, V.~Gluscevic, K.~K. Boddy, E.~O. Nadler, and R.~H. Wechsler,
  ``{Bounds on velocity-dependent dark matter-proton scattering from Milky Way
  satellite abundance},''
  \href{http://dx.doi.org/10.3847/2041-8213/abd807}{{\em Astrophys. J. Lett.}
  {\bfseries 907} no.~2, (2021) L46},
  \href{http://arxiv.org/abs/2010.02936}{{\ttfamily arXiv:2010.02936
  [astro-ph.CO]}}.

\bibitem{Bechtol:2019acd}
K.~Bechtol {\em et~al.}, ``{Dark Matter Science in the Era of LSST},''
  \href{http://arxiv.org/abs/1903.04425}{{\ttfamily arXiv:1903.04425
  [astro-ph.CO]}}.

\bibitem{Ooba:2019erm}
J.~Ooba, H.~Tashiro, and K.~Kadota, ``{Cosmological constraints on the
  velocity-dependent baryon-dark matter coupling},''
  \href{http://dx.doi.org/10.1088/1475-7516/2019/09/020}{{\em JCAP} {\bfseries
  09} (2019) 020}, \href{http://arxiv.org/abs/1902.00826}{{\ttfamily
  arXiv:1902.00826 [astro-ph.CO]}}.

\bibitem{Ali-Haimoud:2015pwa}
Y.~Ali-Ha\"\i{}moud, J.~Chluba, and M.~Kamionkowski, ``{Constraints on Dark
  Matter Interactions with Standard Model Particles from Cosmic Microwave
  Background Spectral Distortions},''
  \href{http://dx.doi.org/10.1103/PhysRevLett.115.071304}{{\em Phys. Rev.
  Lett.} {\bfseries 115} no.~7, (2015) 071304},
  \href{http://arxiv.org/abs/1506.04745}{{\ttfamily arXiv:1506.04745
  [astro-ph.CO]}}.

\bibitem{Ali-Haimoud:2021lka}
Y.~Ali-Ha\"\i{}moud, ``{Testing dark matter interactions with CMB spectral
  distortions},'' \href{http://dx.doi.org/10.1103/PhysRevD.103.043541}{{\em
  Phys. Rev. D} {\bfseries 103} no.~4, (2021) 043541},
  \href{http://arxiv.org/abs/2101.04070}{{\ttfamily arXiv:2101.04070
  [astro-ph.CO]}}.

\bibitem{Rogers:2021byl}
K.~K. Rogers, C.~Dvorkin, and H.~V. Peiris, ``{New limits on light dark matter
  - proton cross section from the cosmic large-scale structure},''
  \href{http://arxiv.org/abs/2111.10386}{{\ttfamily arXiv:2111.10386
  [astro-ph.CO]}}.

\bibitem{Bauer:2020zsj}
J.~B. Bauer, D.~J.~E. Marsh, R.~Hlo\v{z}ek, H.~Padmanabhan, and A.~Lagu\"e,
  ``{Intensity Mapping as a Probe of Axion Dark Matter},''
  \href{http://dx.doi.org/10.1093/mnras/staa3300}{{\em Mon. Not. Roy. Astron.
  Soc.} {\bfseries 500} no.~3, (2020) 3162--3177},
  \href{http://arxiv.org/abs/2003.09655}{{\ttfamily arXiv:2003.09655
  [astro-ph.CO]}}.

\bibitem{Bernal:2020lkd}
J.~L. Bernal, A.~Caputo, and M.~Kamionkowski, ``{Strategies to Detect
  Dark-Matter Decays with Line-Intensity Mapping},''
  \href{http://dx.doi.org/10.1103/PhysRevD.103.063523}{{\em Phys. Rev. D}
  {\bfseries 103} no.~6, (2021) 063523},
  \href{http://arxiv.org/abs/2012.00771}{{\ttfamily arXiv:2012.00771
  [astro-ph.CO]}}.

\bibitem{Das:2017nub}
S.~Das, R.~Mondal, V.~Rentala, and S.~Suresh, ``{On dark matter - dark
  radiation interaction and cosmic reionization},''
  \href{http://dx.doi.org/10.1088/1475-7516/2018/08/045}{{\em JCAP} {\bfseries
  08} (2018) 045}, \href{http://arxiv.org/abs/1712.03976}{{\ttfamily
  arXiv:1712.03976 [astro-ph.CO]}}.

\bibitem{Escudero:2018thh}
M.~Escudero, L.~Lopez-Honorez, O.~Mena, S.~Palomares-Ruiz, and
  P.~Villanueva-Domingo, ``{A fresh look into the interacting dark matter
  scenario},'' \href{http://dx.doi.org/10.1088/1475-7516/2018/06/007}{{\em
  JCAP} {\bfseries 06} (2018) 007},
  \href{http://arxiv.org/abs/1803.08427}{{\ttfamily arXiv:1803.08427
  [astro-ph.CO]}}.

\bibitem{Safarzadeh:2018hhg}
M.~Safarzadeh, E.~Scannapieco, and A.~Babul, ``{A limit on the warm dark matter
  particle mass from the redshifted 21 cm absorption line},''
  \href{http://dx.doi.org/10.3847/2041-8213/aac5e0}{{\em Astrophys. J. Lett.}
  {\bfseries 859} no.~2, (2018) L18},
  \href{http://arxiv.org/abs/1803.08039}{{\ttfamily arXiv:1803.08039
  [astro-ph.CO]}}.

\bibitem{Lopez-Honorez:2018ipk}
L.~Lopez-Honorez, O.~Mena, and P.~Villanueva-Domingo, ``{Dark matter
  microphysics and 21 cm observations},''
  \href{http://dx.doi.org/10.1103/PhysRevD.99.023522}{{\em Phys. Rev. D}
  {\bfseries 99} no.~2, (2019) 023522},
  \href{http://arxiv.org/abs/1811.02716}{{\ttfamily arXiv:1811.02716
  [astro-ph.CO]}}.

\bibitem{Munoz:2019hjh}
J.~B. Mu\~noz, C.~Dvorkin, and F.-Y. Cyr-Racine, ``{Probing the Small-Scale
  Matter Power Spectrum with Large-Scale 21-cm Data},''
  \href{http://dx.doi.org/10.1103/PhysRevD.101.063526}{{\em Phys. Rev. D}
  {\bfseries 101} no.~6, (2020) 063526},
  \href{http://arxiv.org/abs/1911.11144}{{\ttfamily arXiv:1911.11144
  [astro-ph.CO]}}.

\bibitem{Munoz:2020mue}
J.~B. Mu\~noz, S.~Bohr, F.-Y. Cyr-Racine, J.~Zavala, and M.~Vogelsberger,
  ``{ETHOS - an effective theory of structure formation: Impact of dark
  acoustic oscillations on cosmic dawn},''
  \href{http://dx.doi.org/10.1103/PhysRevD.103.043512}{{\em Phys. Rev. D}
  {\bfseries 103} no.~4, (2021) 043512},
  \href{http://arxiv.org/abs/2011.05333}{{\ttfamily arXiv:2011.05333
  [astro-ph.CO]}}.

\bibitem{Jones:2021mrs}
D.~Jones, S.~Palatnick, R.~Chen, A.~Beane, and A.~Lidz, ``{Fuzzy Dark Matter
  and the 21 cm Power Spectrum},''
  \href{http://dx.doi.org/10.3847/1538-4357/abf0a9}{{\em Astrophys. J.}
  {\bfseries 913} no.~1, (2021) 7},
  \href{http://arxiv.org/abs/2101.07177}{{\ttfamily arXiv:2101.07177
  [astro-ph.CO]}}.

\bibitem{Giri:2022nxq}
S.~K. Giri and A.~Schneider, ``{Imprints of fermionic and bosonic mixed dark
  matter on the 21-cm signal at cosmic dawn},''
  \href{http://arxiv.org/abs/2201.02210}{{\ttfamily arXiv:2201.02210
  [astro-ph.CO]}}.

\bibitem{Sarkar:2022dvl}
D.~Sarkar, J.~Flitter, and E.~D. Kovetz, ``{Exploring delaying and heating
  effects on the 21-cm signature of fuzzy dark matter},''
  \href{http://arxiv.org/abs/2201.03355}{{\ttfamily arXiv:2201.03355
  [astro-ph.CO]}}.

\bibitem{Evoli:2014pva}
C.~Evoli, A.~Mesinger, and A.~Ferrara, ``{Unveiling the nature of dark matter
  with high redshift 21 cm line experiments},''
  \href{http://dx.doi.org/10.1088/1475-7516/2014/11/024}{{\em JCAP} {\bfseries
  11} (2014) 024}, \href{http://arxiv.org/abs/1408.1109}{{\ttfamily
  arXiv:1408.1109 [astro-ph.HE]}}.

\bibitem{Fialkov:2018xre}
A.~Fialkov, R.~Barkana, and A.~Cohen, ``{Constraining Baryon--Dark Matter
  Scattering with the Cosmic Dawn 21-cm Signal},''
  \href{http://dx.doi.org/10.1103/PhysRevLett.121.011101}{{\em Phys. Rev.
  Lett.} {\bfseries 121} (2018) 011101},
  \href{http://arxiv.org/abs/1802.10577}{{\ttfamily arXiv:1802.10577
  [astro-ph.CO]}}.

\bibitem{Lidz:2018fqo}
A.~Lidz and L.~Hui, ``{Implications of a prereionization 21-cm absorption
  signal for fuzzy dark matter},''
  \href{http://dx.doi.org/10.1103/PhysRevD.98.023011}{{\em Phys. Rev. D}
  {\bfseries 98} no.~2, (2018) 023011},
  \href{http://arxiv.org/abs/1805.01253}{{\ttfamily arXiv:1805.01253
  [astro-ph.CO]}}.

\bibitem{Munoz:2018pzp}
J.~B. Mu\~noz and A.~Loeb, ``{A small amount of mini-charged dark matter could
  cool the baryons in the early Universe},''
  \href{http://dx.doi.org/10.1038/s41586-018-0151-x}{{\em Nature} {\bfseries
  557} no.~7707, (2018) 684}, \href{http://arxiv.org/abs/1802.10094}{{\ttfamily
  arXiv:1802.10094 [astro-ph.CO]}}.

\bibitem{Munoz:2018jwq}
J.~B. Mu\~noz, C.~Dvorkin, and A.~Loeb, ``{21-cm Fluctuations from Charged Dark
  Matter},'' \href{http://dx.doi.org/10.1103/PhysRevLett.121.121301}{{\em Phys.
  Rev. Lett.} {\bfseries 121} no.~12, (2018) 121301},
  \href{http://arxiv.org/abs/1804.01092}{{\ttfamily arXiv:1804.01092
  [astro-ph.CO]}}.

\bibitem{Kovetz:2018zan}
E.~D. Kovetz, V.~Poulin, V.~Gluscevic, K.~K. Boddy, R.~Barkana, and
  M.~Kamionkowski, ``{Tighter limits on dark matter explanations of the
  anomalous EDGES 21 cm signal},''
  \href{http://dx.doi.org/10.1103/PhysRevD.98.103529}{{\em Phys. Rev. D}
  {\bfseries 98} no.~10, (2018) 103529},
  \href{http://arxiv.org/abs/1807.11482}{{\ttfamily arXiv:1807.11482
  [astro-ph.CO]}}.

\bibitem{Mena:2019nhm}
O.~Mena, S.~Palomares-Ruiz, P.~Villanueva-Domingo, and S.~J. Witte,
  ``{Constraining the primordial black hole abundance with 21-cm cosmology},''
  \href{http://dx.doi.org/10.1103/PhysRevD.100.043540}{{\em Phys. Rev. D}
  {\bfseries 100} no.~4, (2019) 043540},
  \href{http://arxiv.org/abs/1906.07735}{{\ttfamily arXiv:1906.07735
  [astro-ph.CO]}}.

\bibitem{Short:2019twc}
K.~Short, J.~L. Bernal, A.~Raccanelli, L.~Verde, and J.~Chluba, ``{Enlightening
  the dark ages with dark matter},''
  \href{http://dx.doi.org/10.1088/1475-7516/2020/07/020}{{\em JCAP} {\bfseries
  07} (2020) 020}, \href{http://arxiv.org/abs/1912.07409}{{\ttfamily
  arXiv:1912.07409 [astro-ph.CO]}}.

\bibitem{Bernal:2017nec}
J.~L. Bernal, A.~Raccanelli, L.~Verde, and J.~Silk, ``{Signatures of primordial
  black holes as seeds of supermassive black holes},''
  \href{http://dx.doi.org/10.1088/1475-7516/2018/05/017}{{\em JCAP} {\bfseries
  05} (2018) 017}, \href{http://arxiv.org/abs/1712.01311}{{\ttfamily
  arXiv:1712.01311 [astro-ph.CO]}}. [Erratum: JCAP 01, E01 (2020)].

\bibitem{Tashiro:2014tsa}
H.~Tashiro, K.~Kadota, and J.~Silk, ``{Effects of dark matter-baryon scattering
  on redshifted 21 cm signals},''
  \href{http://dx.doi.org/10.1103/PhysRevD.90.083522}{{\em Phys. Rev.}
  {\bfseries D90} no.~8, (2014) 083522},
\href{http://arxiv.org/abs/1408.2571}{{\ttfamily arXiv:1408.2571
  [astro-ph.CO]}}.

\bibitem{Munoz:2015bca}
J.~B. Muoz, E.~D. Kovetz, and Y.~Ali-Hamoud, ``{Heating of Baryons due to
  Scattering with Dark Matter During the Dark Ages},''
  \href{http://dx.doi.org/10.1103/PhysRevD.92.083528}{{\em Phys. Rev.}
  {\bfseries D92} no.~8, (2015) 083528},
\href{http://arxiv.org/abs/1509.00029}{{\ttfamily arXiv:1509.00029
  [astro-ph.CO]}}.

\bibitem{Barkana:2018ab}
R.~Barkana, ``Possible interaction between baryons and dark-matter particles
  revealed by the first stars,''
  \href{http://arxiv.org/abs/1803.06698}{{\ttfamily 1803.06698}}.
  \url{https://arxiv.org/abs/1803.06698}.

\bibitem{Barkana:2018cct}
R.~Barkana, N.~J. Outmezguine, D.~Redigolo, and T.~Volansky, ``{Strong
  constraints on light dark matter interpretation of the EDGES signal},''
  \href{http://dx.doi.org/10.1103/PhysRevD.98.103005}{{\em Phys. Rev.}
  {\bfseries D98} no.~10, (2018) 103005},
\href{http://arxiv.org/abs/1803.03091}{{\ttfamily arXiv:1803.03091 [hep-ph]}}.

\bibitem{Berlin:2018sjs}
A.~Berlin, D.~Hooper, G.~Krnjaic, and S.~D. McDermott, ``{Severely Constraining
  Dark Matter Interpretations of the 21-cm Anomaly},''
  \href{http://dx.doi.org/10.1103/PhysRevLett.121.011102}{{\em Phys. Rev.
  Lett.} {\bfseries 121} no.~1, (2018) 011102},
  \href{http://arxiv.org/abs/1803.02804}{{\ttfamily arXiv:1803.02804
  [hep-ph]}}.

\bibitem{Bowman:2018yin}
J.~D. Bowman, A.~E.~E. Rogers, R.~A. Monsalve, T.~J. Mozdzen, and N.~Mahesh,
  ``{An absorption profile centred at 78 megahertz in the sky-averaged
  spectrum},'' \href{http://dx.doi.org/10.1038/nature25792}{{\em Nature}
  {\bfseries 555} no.~7694, (2018) 67--70},
  \href{http://arxiv.org/abs/1810.05912}{{\ttfamily arXiv:1810.05912
  [astro-ph.CO]}}.

\bibitem{SARAS}
S.~{Singh}, J.~{Nambissan T.}, R.~{Subrahmanyan}, N.~{Udaya Shankar}, B.~S.
  {Girish}, A.~{Raghunathan}, R.~{Somashekar}, K.~S. {Srivani}, and
  M.~{Sathyanarayana Rao}, ``{On the detection of a cosmic dawn signal in the
  radio background},'' \href{http://dx.doi.org/10.1038/s41550-022-01610-5}{{\em
  Nature Astronomy} (Feb., 2022) },
  \href{http://arxiv.org/abs/2112.06778}{{\ttfamily arXiv:2112.06778
  [astro-ph.CO]}}.

\bibitem{LEDA}
D.~C. {Price}, L.~J. {Greenhill}, A.~{Fialkov}, G.~{Bernardi}, H.~{Garsden},
  B.~R. {Barsdell}, J.~{Kocz}, M.~M. {Anderson}, S.~A. {Bourke}, J.~{Craig},
  M.~R. {Dexter}, J.~{Dowell}, M.~W. {Eastwood}, T.~{Eftekhari}, S.~W.
  {Ellingson}, G.~{Hallinan}, J.~M. {Hartman}, R.~{Kimberk}, T.~J.~W. {Lazio},
  S.~{Leiker}, D.~{MacMahon}, R.~{Monroe}, F.~{Schinzel}, G.~B. {Taylor},
  E.~{Tong}, D.~{Werthimer}, and D.~P. {Woody}, ``{Design and characterization
  of the Large-aperture Experiment to Detect the Dark Age (LEDA) radiometer
  systems},'' \href{http://dx.doi.org/10.1093/mnras/sty1244}{{\em Mon. Not.
  Roy. Astron. Soc.} {\bfseries 478} no.~3, (Aug., 2018) 4193--4213},
  \href{http://arxiv.org/abs/1709.09313}{{\ttfamily arXiv:1709.09313
  [astro-ph.IM]}}.

\bibitem{DeBoer:2016tnn}
D.~R. DeBoer {\em et~al.}, ``{Hydrogen Epoch of Reionization Array (HERA)},''
  \href{http://dx.doi.org/10.1088/1538-3873/129/974/045001}{{\em Publ. Astron.
  Soc. Pac.} {\bfseries 129} no.~974, (2017) 045001},
  \href{http://arxiv.org/abs/1606.07473}{{\ttfamily arXiv:1606.07473
  [astro-ph.IM]}}.

\bibitem{SKA:2018ckk}
{\bfseries SKA} Collaboration, D.~J. Bacon {\em et~al.}, ``{Cosmology with
  Phase 1 of the Square Kilometre Array: Red Book 2018: Technical
  specifications and performance forecasts},''
  \href{http://dx.doi.org/10.1017/pasa.2019.51}{{\em Publ. Astron. Soc.
  Austral.} {\bfseries 37} (2020) e007},
  \href{http://arxiv.org/abs/1811.02743}{{\ttfamily arXiv:1811.02743
  [astro-ph.CO]}}.

\bibitem{Lewis:2007zh}
A.~Lewis, ``{Linear effects of perturbed recombination},''
  \href{http://dx.doi.org/10.1103/PhysRevD.76.063001}{{\em Phys. Rev. D}
  {\bfseries 76} (2007) 063001},
  \href{http://arxiv.org/abs/0707.2727}{{\ttfamily arXiv:0707.2727
  [astro-ph]}}.

\bibitem{Blas:2011rf}
D.~Blas, J.~Lesgourgues, and T.~Tram, ``{The Cosmic Linear Anisotropy Solving
  System (CLASS) II: Approximation schemes},''
  \href{http://dx.doi.org/10.1088/1475-7516/2011/07/034}{{\em JCAP} {\bfseries
  07} (2011) 034}, \href{http://arxiv.org/abs/1104.2933}{{\ttfamily
  arXiv:1104.2933 [astro-ph.CO]}}.

\bibitem{Naoz:2005pd}
S.~Naoz and R.~Barkana, ``{Growth of linear perturbations before the era of the
  first galaxies},''
  \href{http://dx.doi.org/10.1111/j.1365-2966.2005.09385.x}{{\em Mon. Not. Roy.
  Astron. Soc.} {\bfseries 362} (2005) 1047--1053},
  \href{http://arxiv.org/abs/astro-ph/0503196}{{\ttfamily
  arXiv:astro-ph/0503196}}.

\bibitem{Tseliakhovich:2010yw}
D.~Tseliakhovich, R.~Barkana, and C.~Hirata, ``{Suppression and Spatial
  Variation of Early Galaxies and Minihalos},''
  \href{http://dx.doi.org/10.1111/j.1365-2966.2011.19541.x}{{\em Mon. Not. Roy.
  Astron. Soc.} {\bfseries 418} (2011) 906},
  \href{http://arxiv.org/abs/1012.2574}{{\ttfamily arXiv:1012.2574
  [astro-ph.CO]}}.

\bibitem{Tseliakhovich:2010bj}
D.~Tseliakhovich and C.~Hirata, ``{Relative velocity of dark matter and
  baryonic fluids and the formation of the first structures},''
  \href{http://dx.doi.org/10.1103/PhysRevD.82.083520}{{\em Phys. Rev. D}
  {\bfseries 82} (2010) 083520},
  \href{http://arxiv.org/abs/1005.2416}{{\ttfamily arXiv:1005.2416
  [astro-ph.CO]}}.

\bibitem{Dalal:2010yt}
N.~Dalal, U.-L. Pen, and U.~Seljak, ``{Large-scale BAO signatures of the
  smallest galaxies},''
  \href{http://dx.doi.org/10.1088/1475-7516/2010/11/007}{{\em JCAP} {\bfseries
  11} (2010) 007}, \href{http://arxiv.org/abs/1009.4704}{{\ttfamily
  arXiv:1009.4704 [astro-ph.CO]}}.

\bibitem{Fialkov:2011iw}
A.~Fialkov, R.~Barkana, D.~Tseliakhovich, and C.~M. Hirata, ``{Impact of the
  Relative Motion between the Dark Matter and Baryons on the First Stars},''
  \href{http://dx.doi.org/10.1111/j.1365-2966.2012.21318.x}{{\em Mon. Not. Roy.
  Astron. Soc.} {\bfseries 424} (2012) 1335--1345},
  \href{http://arxiv.org/abs/1110.2111}{{\ttfamily arXiv:1110.2111
  [astro-ph.CO]}}.

\bibitem{Senatore:2008vi}
L.~Senatore, S.~Tassev, and M.~Zaldarriaga, ``{Cosmological Perturbations at
  Second Order and Recombination Perturbed},''
  \href{http://dx.doi.org/10.1088/1475-7516/2009/08/031}{{\em JCAP} {\bfseries
  08} (2009) 031}, \href{http://arxiv.org/abs/0812.3652}{{\ttfamily
  arXiv:0812.3652 [astro-ph]}}.

\bibitem{Seager:1999bc}
S.~Seager, D.~D. Sasselov, and D.~Scott, ``{A new calculation of the
  recombination epoch},'' \href{http://dx.doi.org/10.1086/312250}{{\em
  Astrophys. J. Lett.} {\bfseries 523} (1999) L1--L5},
  \href{http://arxiv.org/abs/astro-ph/9909275}{{\ttfamily
  arXiv:astro-ph/9909275}}.

\bibitem{Rubino-Martin:2009frf}
J.~A. Rubino-Martin, J.~Chluba, W.~A. Fendt, and B.~D. Wandelt, ``{Estimating
  the impact of recombination uncertainties on the cosmological parameter
  constraints from cosmic microwave background experiments},''
  \href{http://dx.doi.org/10.1111/j.1365-2966.2009.16136.x}{{\em Mon. Not. Roy.
  Astron. Soc.} {\bfseries 403} (2010) 439},
  \href{http://arxiv.org/abs/0910.4383}{{\ttfamily arXiv:0910.4383
  [astro-ph.CO]}}.

\bibitem{Lee:2020obi}
N.~Lee and Y.~Ali-Ha\"\i{}moud, ``{HYREC-2: a highly accurate sub-millisecond
  recombination code},''
  \href{http://dx.doi.org/10.1103/PhysRevD.102.083517}{{\em Phys. Rev. D}
  {\bfseries 102} no.~8, (2020) 083517},
  \href{http://arxiv.org/abs/2007.14114}{{\ttfamily arXiv:2007.14114
  [astro-ph.CO]}}.

\bibitem{Ali-Haimoud:2013hpa}
Y.~Ali-Ha\"\i{}moud, P.~D. Meerburg, and S.~Yuan, ``{New light on 21 cm
  intensity fluctuations from the dark ages},''
  \href{http://dx.doi.org/10.1103/PhysRevD.89.083506}{{\em Phys. Rev. D}
  {\bfseries 89} no.~8, (2014) 083506},
  \href{http://arxiv.org/abs/1312.4948}{{\ttfamily arXiv:1312.4948
  [astro-ph.CO]}}.

\bibitem{Lewis:2007kz}
A.~Lewis and A.~Challinor, ``{The 21cm angular-power spectrum from the dark
  ages},'' \href{http://dx.doi.org/10.1103/PhysRevD.76.083005}{{\em Phys. Rev.
  D} {\bfseries 76} (2007) 083005},
  \href{http://arxiv.org/abs/astro-ph/0702600}{{\ttfamily
  arXiv:astro-ph/0702600}}.

\bibitem{Pillepich:2006fj}
A.~Pillepich, C.~Porciani, and S.~Matarrese, ``{The bispectrum of redshifted
  21-cm fluctuations from the dark ages},''
  \href{http://dx.doi.org/10.1086/517963}{{\em Astrophys. J.} {\bfseries 662}
  (2007) 1--14},
\href{http://arxiv.org/abs/astro-ph/0611126}{{\ttfamily arXiv:astro-ph/0611126
  [astro-ph]}}.

\bibitem{driskell_in_prep}
T.~{Driskell} {\em et~al.}, ``{Bounds on dark matter-baryon scattering from the
  global 21~cm signal and the suppressed matter power spectrum},'' {\em In
  prep.} (2022) .

\end{thebibliography}\endgroup
\end{document}